\documentclass[fleqn,usenatbib]{mnras}

\usepackage[T1]{fontenc}

\DeclareRobustCommand{\VAN}[3]{#2}
\let\VANthebibliography\thebibliography
\def\thebibliography{\DeclareRobustCommand{\VAN}[3]{##3}\VANthebibliography}

\usepackage{graphicx}	
\usepackage{amsmath}	
\usepackage{amssymb}	

\title[Impact of the CMB on metal-poor IMF]
{Impact of the cosmic background radiation on the initial mass function of metal-poor stars}

\author[S. Chon, H. Ono, K. Omukai, \& R. Schneider]{
Sunmyon Chon $^{1}$\thanks{E-mail: sunmyon.chon@astr.tohoku.ac.jp},
Haruka Ono $^{1}$,
Kazuyuki Omukai $^{1}$, and
Raffaella Schneider $^{2,3,4}$
\\
$^{1}$Astronomical Institute, Graduate School of Science, Tohoku University, Aoba, Sendai 980-8578, Japan\\
$^{2}$Dipartimento di Fisica, Universit\`{a} di Roma ‘La Sapienza’, P.le Aldo Moro 2, I-00185 Roma, Italy \\
$^{3}$INAF/Osservatorio Astronomico di Roma, via di Frascati 33, I-00078 Monteporzio Catone, Italy \\
$^{4}$INFN, Sezione di Roma 1, P.le Aldo Moro 2, I-00185 Roma, Italy
}

\date{Accepted XXX. Received YYY; in original form ZZZ}

\pubyear{2022}

\begin{document}
\label{firstpage}
\pagerange{\pageref{firstpage}--\pageref{lastpage}}
\maketitle

\begin{abstract}
We study star cluster formation at low metallicities of $Z/Z_\odot=10^{-4}$--$10^{-1}$ using three-dimensional hydrodynamics simulations.
Particular emphasis is put on how the stellar mass distribution is affected by the cosmic microwave background radiation (CMB), which sets the temperature floor to the gas.
Starting from the collapse of a turbulent cloud, we follow the formation of a protostellar system resolving $\sim$au scale.
In relatively metal-enriched cases of $Z/Z_\odot \gtrsim 10^{-2}$, where the mass function resembles the present-day one in the absence of the CMB,  
high temperature CMB suppresses cloud fragmentation and reduces the number of low-mass stars, making the mass function more top-heavy than in the cases without CMB heating at $z\gtrsim10$.
In lower-metallicity cases with $Z/Z_\odot \lesssim 10^{-3}$, where the gas temperature is higher than the CMB value due to inefficient cooling, the CMB has only a minor impact on the mass distribution, which is top-heavy regardless of the redshift.
In cases either with a low metallicity of $Z/Z_\odot \lesssim 10^{-2}$ or at a high redshift $z\gtrsim10$,
the mass spectrum consists of a low-mass Salpeter-like component, peaking at $0.1~M_\odot$, and a top-heavy component with $10$--$50~M_\odot$, with the fraction in the latter increasing with increasing redshift.
In galaxies forming at $z\gtrsim10$, the major targets of the future instruments including JWST,  
CMB heating makes the stellar mass function significantly top-heavy, enhancing the number of supernova explosions by a factor of $1.4$ ($2.8$) at $z=10$ ($20$, respectively) compared to the prediction by Chabrier initial mass function when $Z/Z_\odot=0.1$.
\end{abstract}



\section{Introduction}
The initial mass function (IMF) of stars formed in the metal-poor environments characterizing high-redshift galaxies played a crucial role in shaping the present-day Universe.
Star formation has impact on the evolution of the interstellar and inter-galactic media (ISM and IGM, respectively) via its feedback processes, with their strength depending strongly on the stellar masses \citep[see e.g.,][]{Ciardi+2005}.
In particular, massive stars alter their environments drastically, for example by emitting ultraviolet (UV) radiation and by ionizing and heating up the surrounding gas.
Stars more massive than $\sim 8~M_\odot$ end their lives with violent supernova (SN) explosions, injecting a large amount of kinetic energy into the ISM, as well as materials synthesized in the stellar interior, enriching the Universe with heavy elements.
More massive stars with masses $> 30$--$40~M_{\odot}$ may fail to explode and collapse directly to black holes (BHs), 
which may eventually grow to become the supermassive BHs residing at the centers of galaxies 
\citep[e.g.][]{Volonteri2010, Haiman2013, Chon+2016, Valiante+2016, Valiante+2018, Pezzulli+2017,Inayoshi+2020, Sessano+2021}.
Accreting BHs also exert strong feedback on the surrounding gas by emitting X-rays and may play some role in cosmic reionization \citep[][]{Alvarez+2009, Smith+2011, Jeon+2014, Aykutalp+2014, Graziani+2018, Dayal+2020, Chon+2021a}.

Our knowledge of the stellar IMF in metal-poor environments is, however, still very limited. 
Numerical simulations predict that the first stars formed from primordial pristine gas are typically much more massive, $\sim 100~M_\odot$
\citep[][]{OmukaiNishi1998, Abel+2002, Bromm2002, OmukaiPalla2003, Yoshida+2008,
Hosokawa+2011, Hosokawa+2016, Hirano+2014, Hirano+2015, Susa+2014, Sugimura+2020},
than stars in the present-day universe, $0.1$--$1~M_\odot$
\citep[e.g.][]{Kroupa2002, Chabrier2003} although a small number of lower-mass stars are also expected to form
\citep[][]{Machida+2008b, Clark+2011, Smith+2011, Greif+2012, Stacy+2016, Susa2019, Latif+2022}.
Observations of low-mass metal-poor stars still surviving in the present-day universe, i.e., in the Galactic halo or local dwarf galaxies, suggest that only a small number of low-mass stars were formed in the early universe, i.e., that their IMF was top-heavy
\citep[][]{Salvadori+2007, Salvadori+2008, Bennassuti+2014, Bennassuti+2017, Graziani+2015, Graziani+2017, Hartwig+2015, Ishiyama+2016, Hartwig+2018b, Magg+2018}.
This suggests that the transition from top-heavy to bottom-heavy Salpeter-like IMF occurred at some point in the history of the universe. 
What drives the stellar mass transition and how it proceeds in the course of galaxy formation still remains largely unanswered.

One school of thought tried to explain the IMF from the view point of mass scales of gravitational fragmentation, 
which may be related to thermal properties of star-forming gas
\citep[e.g.][]{Larson1985, Larson1998, Larson2005, Tsuribe&Omukai2006, Bonnell+2006}.
With efficient cooling, clouds are distorted in shape and become highly filamentary in the course of gravitational contraction.
Such filaments then fragment, producing a number of protostars.
This mode of fragmentation occurs efficiently when the effective specific heat ratio of the gas $\gamma$ is smaller than unity,
where $\gamma \equiv \mathrm{d} \log P / \mathrm{d} \log \rho$ and 
$P$ and $\rho$ are the pressure and density of the gas, respectively 
\citep[][]{Larson1985, Inutsuka&Miyama1992, Li+2003, Jappsen+2005, Sugimura+2017}.
In the course of collapse, the filamentary structure stops developing when $\gamma$ becomes larger than unity and further fragmentation is suppressed after this moment, thereby setting the minimum fragmentation scale of the cloud.

Thermal properties of star-forming gas are largely controlled by its metallicity. 
Metals, which can be in the gas phase or in dust grains, efficiently remove the thermal energy of the gas via line emission or dust thermal emission \citep[e.g.][]{Omukai2000, Bromm+2001b, Bromm&Loeb2003, Schneider+2003, Schneider+2006, Schneider+2012, Omukai+2005,Smith+2009, Schneider&Omukai2010, Chiaki+2014}.
With more metals, the gas temperature and thus the Jeans or fragmentation mass scale becomes smaller.
In this way, metal enrichment in the ISM lowers the typical mass of forming stars and drives the transition from the primordial top-heavy to the present-day bottom-heavy IMF.
In fact, numerical simulations have demonstrated that
a cloud with a trace amount of metals fragments into a number of low-mass stars with mass $0.01$--$1~M_\odot$, owing to dust cooling \citep{Tsuribe&Omukai2006, Tsuribe&Omukai2008, Clark+2008, Jappsen+2009, Dopcke+2013, Safranek-Shrader+2014, Safranek-Shrader+2016, Chiaki+2016, Chiaki+2021, Chon+2021b, Shima&Hosokawa2021}.

Stellar mass is, however, determined not only by cloud fragmentation but also by subsequent protostellar accretion. 
Forming inside dense cores produced by fragmentation, protostars can gain mass by accreting the matter not only from inside but also from outside their parental cores.
Protostellar accretion continues for $10^4$ -- $10^5~$yr, until the accretion flow is shut off by stellar radiative feedback \citep[e.g.][]{Peters+2010, Geen+2018, He+2019, Fukushima+2020b}.
This means that to understand the origin of the stellar IMF,
not only cloud fragmentation and birth of the protostellar cores, but also long-term accretion onto the protostars need to be investigated.
Some authors in fact claim that mass accretion is the key process in shaping the IMF, with more massive stars accreting a larger amount of gas, building up the Salpeter-like power-law stellar mass spectrum as the outcome \citep[e.g.][]{Bonnell+2001, Bate+2003}. 

Recently, \citet[][hereafter Paper~I]{Chon+2021b} investigated star cluster formation for a wide range of metallicities $10^{-6} < Z/Z_\odot < 10^{-1}$ 
by following the entire evolution starting from an initially turbulent cloud up to $10^4$--$10^5~$years of protostellar accretion.  
They found that the mass distribution of forming protostars is top-heavy at very low metallicities but approaches the present-day Salpeter-like IMF when the metallicity reaches $Z/Z_\odot = 0.01$--$0.1$.
In those calculations, the mass distribution is determined by the interplay between turbulence and cooling-induced fragmentation.
When the metallicity is as high as $Z/Z_\odot = 0.1$, the initial turbulent motion creates a filamentary structure, which then fragments into a number of protostars due to efficient cooling, and the mass reservoir in the initial cloud is shared among them, leading to a bottom-heavy Salpeter-like mass distribution.
On the other hand, at lower metallicities ($Z/Z_\odot \lesssim 0.01-0.1$), turbulence quickly decays and a massive core forms at the cloud center.
A small number of stars in the central region accrete most of the mass, leading to a top-heavy mass distribution.
Although a number of low-mass stars are simultaneously formed in the peripherical regions, most of them accrete only a small fraction of the total stellar mass and fail to grow massive. 

Another factor that needs to be considered when studying star formation in the early universe is the presence of the cosmic microwave background radiation (CMB). 
Heating by the CMB has potential impact on the fragmentation scale of the clouds by setting a redshift-dependent minimum temperature floor of $T_\text{CMB} = 2.73(1 + z)~$K to which the gas is able to cool. 
In particular, a gas with modest metal-enrichment at $Z/Z_\odot \gtrsim 10^{-3}$--$10^{-2}$ contracts almost isothermally at the CMB temperature during the prestellar phase \citep[e.g.][]{Bromm+2001b, Omukai+2005, Smith+2009, Schneider&Omukai2010}.
Such isothermal ($\gamma=1$) evolution may inhibit cloud fragmentation, compared to the case with more rapid cooling ($\gamma<1$), resulting in a larger fragmentation mass scale \citep{Schneider+2012, Riaz+2020}.
Also protostellar accretion evolution may be modified due to higher gas temperature.

In this paper, we investigate how the difference in CMB temperature affects the mass distribution of forming stars, by following long-term accretion evolution of protostars as in Paper~I.
This paper is organized as follows.
We describe the initial condition and the numerical methodology in Section~\ref{sec::method}.
We present our numerical results in Section~\ref{sec::results} and discuss the implication of our results in Section~\ref{sec::discussion}.
We summarize our findings in Section~\ref{sec::conclusion}.

\section{Methodology} \label{sec::method}
We perform hydrodynamics simulations using the smoothed particle hydrodynamic (SPH) code, {\tt GADGET-2} \citep{Springel2005}
to follow the formation of stellar systems in metal-poor environments in the early universe.
We consider combinations of four different metallicities ($\log Z/Z_\odot = -1$, $-2$, $-3$, and $-4$)
and four different redshifts ($z = 0$, $5$, $10$, and $20$), i.e., 16 models in total.
Starting from the collapse of an initially turbulent cloud, 
we follow the formation of a multiple protostellar system.
The external CMB sets the temperature floor to the gas and dust by heating them if their temperatures are below the CMB temperature.
In this section, we mainly describe the numerical implementation of the CMB effect.
Since the numerical methodology other than the CMB part has been thoroughly described in Paper~I, where we investigated metallicity effects, we just briefly overview it. 

As our initial condition, we select a critical Bonnor-Ebert sphere with a central density of $10^3~\mathrm{cm^{-3}}$ and a temperature of $200~$K, and enhance the density by $1.4$ times to 
trigger the gravitational collapse. 
The cloud mass and radius are $6300~M_\odot$ and $5 \times 10^6~$au (25 pc), respectively.
We also impose rigid rotation and turbulent motion to construct the initial velocity field.
The turbulent velocity field is generated as in \citet{MacLow1999} with the power spectrum $P(k) \propto k^{-4}$ \citep[e.g.][]{Larson1985}.
The amplitude of the turbulence is chosen to be $\mathcal{M}_\text{ch} \equiv v_\text{disp} / c_\text{s} = 1$,
where $v_\text{disp}$ is the mass-weighted root-mean square of the random velocity field,
and $c_\text{s}$ is the sound speed.
We also impose rigid rotation with energy $10^{-2}$ of the gravitational energy.

To ensure sufficient resolution to follow the gravitational collapse,
we perform particle splitting when the gas density exceeds a threshold value $n_\text{split}$.
A single SPH particle is split into 13 daughter particles following the prescription by  \citet{Kitsionas+2002}.
At each time, particle splitting is performed at two threshold densities $n_\text{split} = 10^5$ and $10^{8}~\mathrm{cm^{-3}}$ 
before the formation of protostars.
The particle mass is $0.016~M_\odot$ before the splitting and 
is $3.6 \times 10^{-4}$ ($2.8 \times 10^{-5}~M_\odot$) after the first (second, respectively) splitting. 
This allows us to resolve the local Jeans mass $M_\text{J}$ with more than $100$--$1000$ SPH particles, adequate to follow the gravitational collapse of a cloud \citep{Bate+1995, Truelove1997}.
Note that we choose $n_\text{split}$ so as not to execute the particle splitting in circumstellar disks,
which would induce numerical density perturbations and trigger spurious fragmentation \citep{Chiaki+2016}.
When the gas density reaches $n_\text{sink} = 2 \times 10^{15}~\mathrm{cm^{-3}}$, 
we introduce a sink particle.
We set the size of the sink region to be $\sim 1~$au and assume the gas particles entering it are assimilated to the sink particle.

\subsection{chemistry and thermal processes}
We calculate non-equilibrium chemistry of eight primordial species, e$^-$, H, H$^+$, H$^-$, H$_2$, D, D$^+$, and HD, 
with 22 chemical reactions among them.
For the C- and O-bearing species, we just assume that all the C and O are in the form of C~{\sc ii} and O~{\sc i} without solving their chemical reactions. 
This prescription allows us to reproduce approximately the thermal evolution 
with detailed chemical network for C- and O-bearing coolants such as CO, OH, and H$_2$O \citep{Omukai+2005}.
The assumed elemental number fractions of C and O nuclei are 
$y_\text{C, gas} = 9.27 \times 10^{-3}$ and $y_\text{O, gas} = 3.568 \times 10^{-3}$ with respect to H nuclei at $Z=Z_{\odot}$ \citep{Omukai+2005}.
We scale the abundances of heavy elements in proportion to the metallicity at lower metallicities. 

In addition to the chemical cooling/heating processes, we consider radiative cooling by the following processes; line cooling by H, H$_2$, HD, and fine-structure lines of C~{\sc ii} and O~{\sc i}
and continuum cooling by the primordial gas \citep{Matsukoba+2019}.
We also consider dust thermal emission with the assumption that the size distribution and the chemical composition of dust grains follow those observed in the Milky Way \citep{Semenov+2003}.

\begin{figure*}
	\centering
		\includegraphics[width=16.5cm]{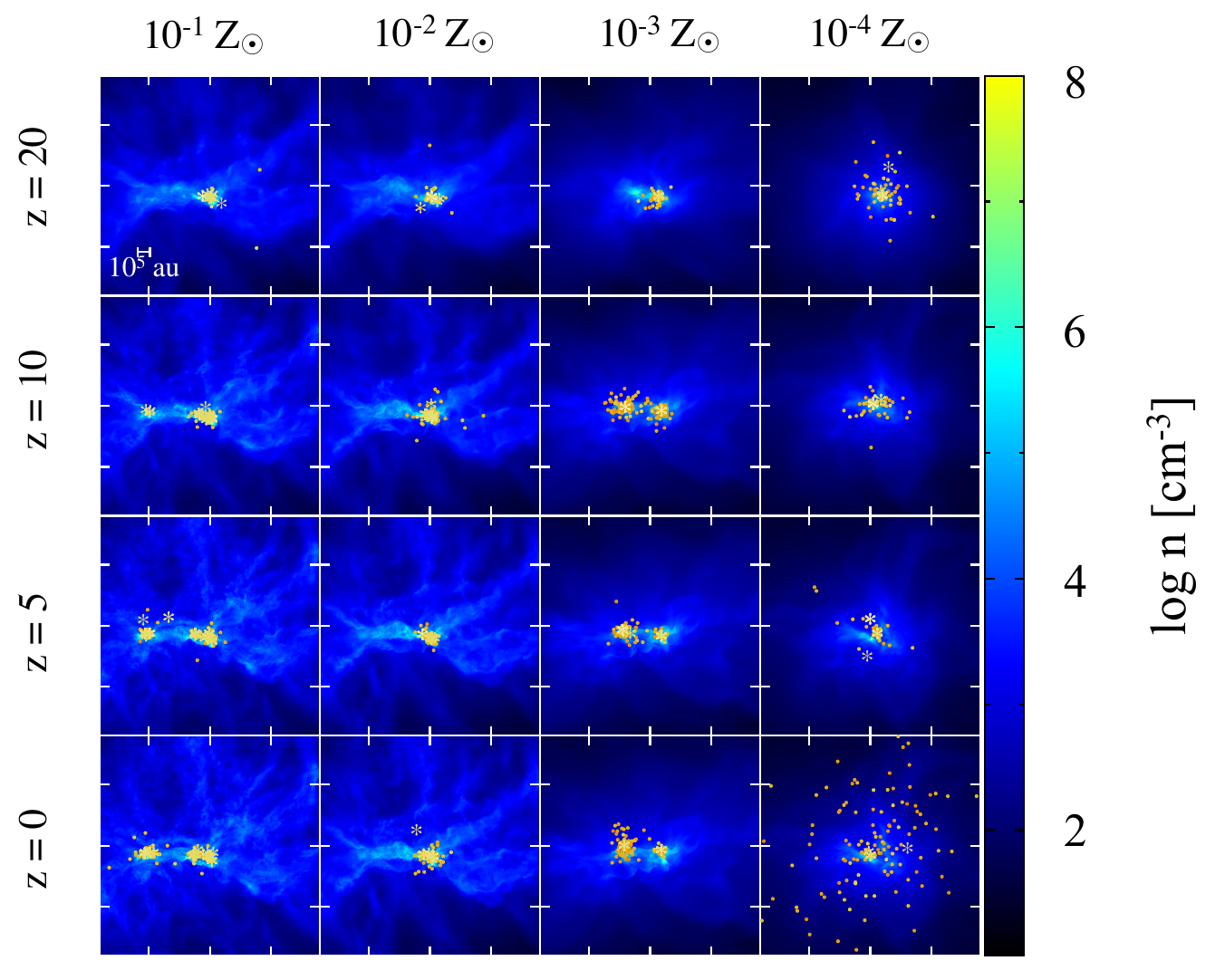}
		\caption{The projected density distributions
		for cases with different metallicities ($Z/Z_{\odot}=10^{-1}, 10^{-2}, 10^{-3}, 10^{-4}$: columns) and redshifts ($z=0, 5, 10, 20$: rows) when the first protostar forms at the density $2\times 10^{15}~\mathrm{cm^{-3}}$.
		We overplot the positions of protostars at a later time 
		when the total stellar mass reaches $150~M_\odot$ on top panels by the symbols: the asterisks (dots) show the stars with the masses larger (smaller, respectively) than $1~M_\odot$.
		}
		\label{fig_snapshot}
\end{figure*}

\begin{figure*}
	\centering
		\includegraphics[width=16.5cm]{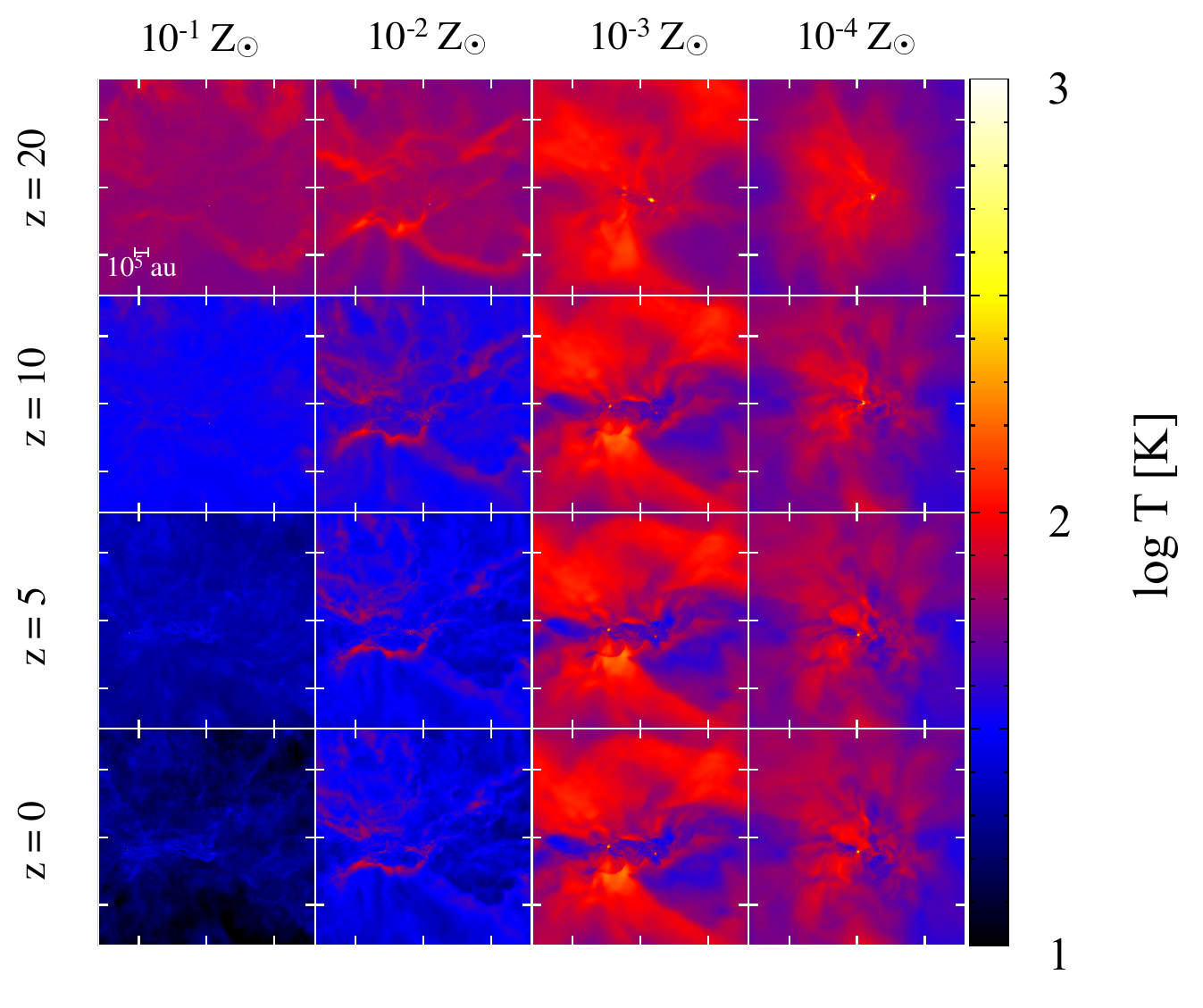}
		\caption{The projected temperature distributions 
		for cases with different metallicities ($Z/Z_{\odot}=10^{-1}, 10^{-2}, 10^{-3}, 10^{-4}$: columns) and redshifts ($z=0, 5, 10, 20$: rows) when the first protostar forms at the density $2\times 10^{15}~\mathrm{cm^{-3}}$.
		}
		\label{fig_snapshot_Tgas}
\end{figure*}

\subsection{CMB heating}
Rather than setting the minimum temperature by hand, we include the effect of CMB heating in the following way.
The CMB effect enters in the line cooling rate by modifying the level population.  
We treat C~{\sc ii} and O~{\sc i} as two level systems, 
considering the detailed balance between the levels $R_{21}n_{2}=R_{12}n_{1}$, 
where $n_i$ is the occupation number of level $i$ and  $R_{ij}$ is the transition rate from level $i$ to $j$.
The transition rates can be described as \citep{Tielens+1985, Omukai2001},
\begin{align}
R_{21} &= A_{21} (1 + Q_{21})  + C_{21}, \\
R_{12} &= g_{2}/g_{1} A_{21} Q_{21} + C_{12}, \\
Q_{21} &= \left\{ \exp \left( \frac{h\nu_{21}}{k_\text{B}T_\text{CMB}} \right) - 1 \right\}^{-1},
\end{align}
where $A_{21}$ is the spontaneous transition probability, $g_{i}$ is the statistical weight of level $i$,
and $C_{ij}$ is the collisional excitation/deexcitation rates from level $i$ to $j$.
The term $Q_{21}$ describes the level pumping by the external CMB radiation,
where $h\nu_{21}$ is the energy difference between the levels $1$ and $2$ and $k_\text{B}$ is the Boltzmann constant.
The line cooling rate $\Lambda_\text{line}$ becomes
\begin{align}
\Lambda_\text{line} = h\nu_{21} A_{21} n_{2} \left \{1 -  Q_{21} \left( \frac{g_2n_1}{g_1n_2} - 1\right) \right \}.
\end{align}
As for the HD line cooling, we use the fitting function 
in \citet{Flower+2000}, which also includes the CMB effect.
For H and H$_2$ lines,  
since the energy level difference $h\nu_{21}/k_\text{B}$ 
is far larger than the CMB temperature, 
the CMB radiation has negligible impact on the level population, and we do not consider the CMB effect.

The CMB effect on the dust cooling rate enters by modifying the dust temperature 
$T_\text{dust}$, which is determined by the energy balance of the dust grains,
\begin{align} \label{eq::Tdust}
4 \sigma T_\text{dust}^4 \kappa_\text{gr} \rho= \Lambda_\mathrm{gas\rightarrow dust} + 4\sigma T_\text{CMB}^4 \kappa_\text{gr} \rho,
\end{align}
where $\sigma$ is the Stefan-Boltzmann constant, 
$\kappa_\text{gr}$ is the absorption opacity of dust grains,
$\Lambda_\mathrm{gas\rightarrow dust}$ is the rate of the energy transferred to the gas 
due to the collision between gas and dust grains given by \citet{HM1979}.

\begin{figure*}
	\centering
		\includegraphics[width=18.cm]{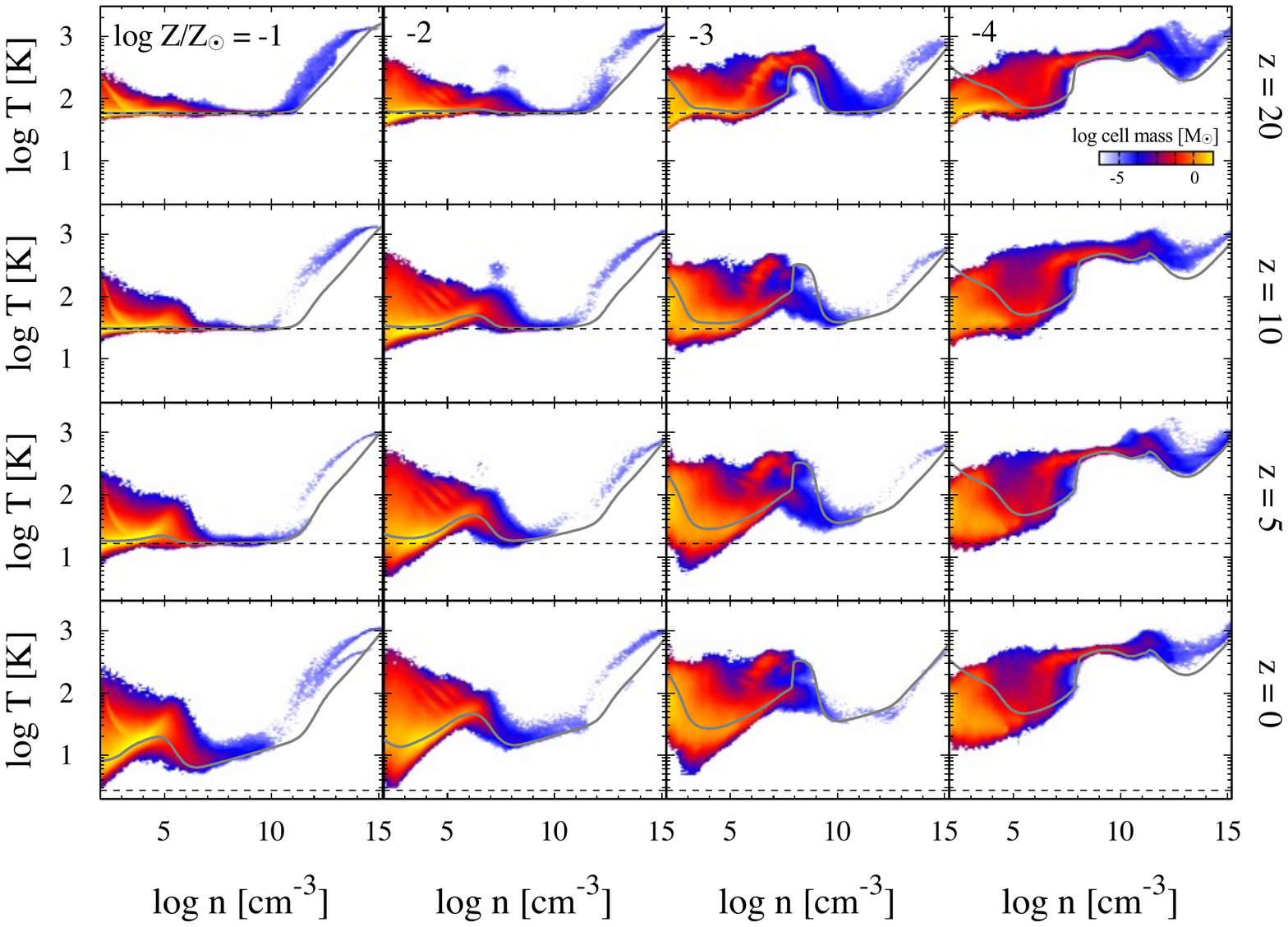}
		\caption{The number density ($n$) versus temperature ($T$) diagrams for different metallicity and redshift runs.
		We divide each $n$-$T$ plane into $200\times200$ cells,
		where the colors show the gas mass in each cells.
		The grey lines show the temperature evolution calculated by the one-zone model.
		The dashed lines represent the CMB temperature.}
		\label{fig_rhoT_hist}
\end{figure*}

\begin{figure}
	\centering
		\includegraphics[width=8.cm]{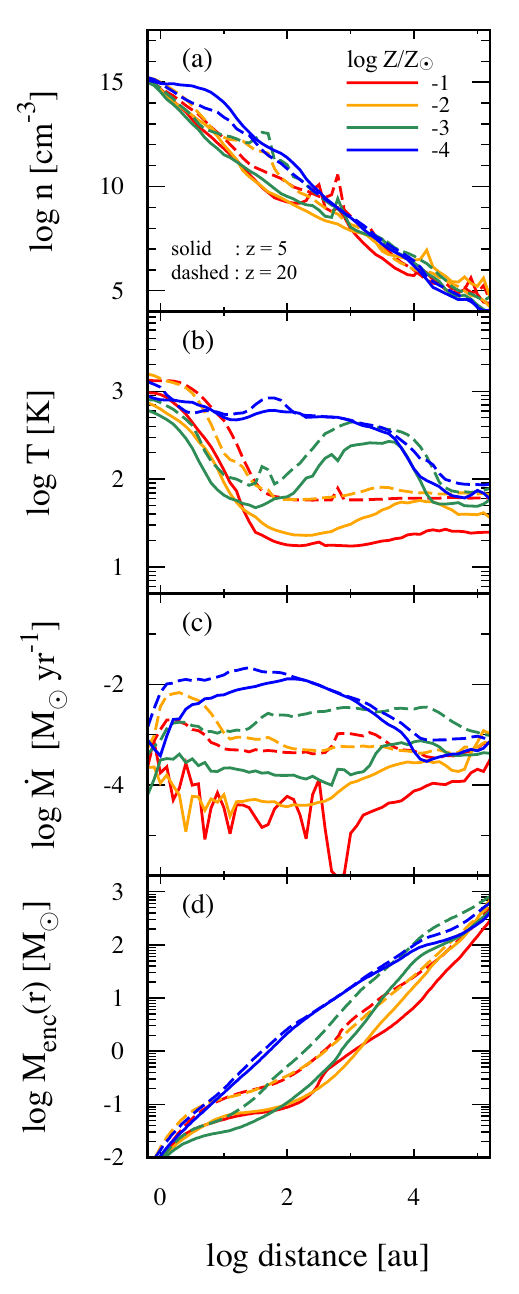}
		\caption{
		The radial profiles of (a) gas density $n$, (b) temperature $T$, (c) mass infall rate $\dot{M}_\text{inf}$,
		and (d) enclosed mass $M_\text{enc} (r)$ as a function of the distance from the cloud center.
		The profile is taken at the moment when the central cloud density reaches $2\times 10^{15}~\mathrm{cm^{-3}}$,
		when the first protostar forms.
		The line colors indicate the cases with $\log Z/Z_\odot = -1$ (red), $-2$ (orange), $-3$ (green), and $-4$ (blue).
        Solid (dashed) lines show the cases with $z=5$ ($20$).
		}
		\label{fig_radial_profiles}
\end{figure}

\begin{figure}
	\centering
		\includegraphics[width=7.5cm]{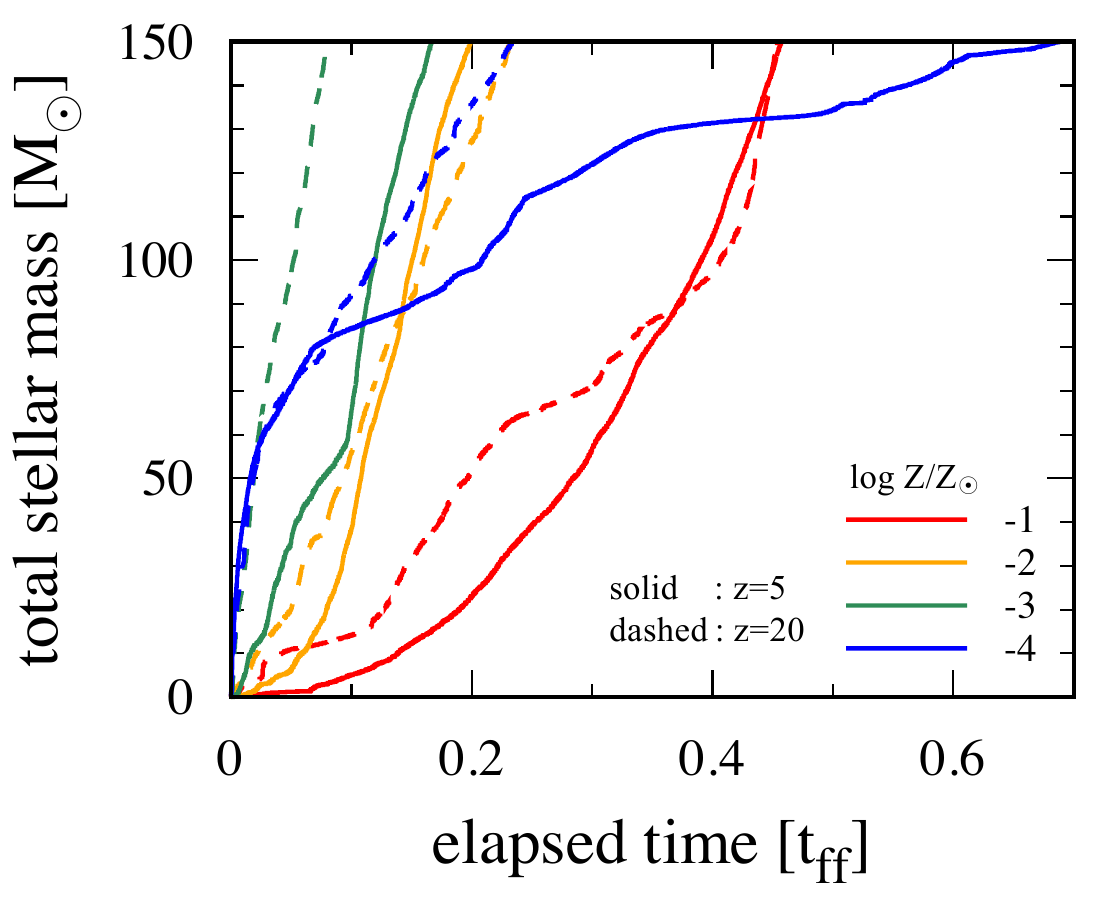}
		\caption{
		The time evolution of the total stellar mass for the cases with different metallicities $\log Z / Z_\odot = -1$ (red), $-2$ (orange), $-3$ (green), and $-4$ (blue) at two different epochs $z=5$ (solid) and $z=20$ (dashed). 
		We normalize the elapsed time by the free-fall time at the initial cloud density $n=10^3~\mathrm{cm^{-3}}$, $t_\text{ff} = 5.4 \times 10^5~$yr. 
		}
		\label{fig_mtot_evolution}
\end{figure}

\section{Result} \label{sec::results}
In this section, we describe how different temperatures of the CMB affect the cloud evolution and mass distribution of forming stars.
In section~\ref{sec::collapse}, we discuss the evolution until the formation of the first protostar.
We focus on how cloud properties, e.g., thermal state and cloud morphology, change with different CMB temperatures. 
Note that the cloud morphology is crucial for determining the shape of the mass spectra, as seen in Paper~I. 
In section~\ref{sec::stellar_system}, 
we investigate the emergence of multiple stellar systems.
In section~\ref{sec::spectrum}, 
we present the mass spectra for different values of the CMB temperature and analyze their redshift evolution.

\subsection{Prestellar evolution: cloud collapse and fragmentation} \label{sec::collapse}
Fig.~\ref{fig_snapshot} (\ref{fig_snapshot_Tgas}) shows the projected density (temperature) maps at the time of the first protostar formation for the cases with different metallicities ($Z/Z_{\odot}=10^{-1}, 10^{-2}, 10^{-3}$, and $10^{-4}$) at four different redshifts ($z=0, 5, 10$, and $20$).
The temperature versus density diagrams at the same epochs are shown in Fig.~\ref{fig_rhoT_hist}, where the temperature-density plane is divided into $200 \times 200$ cells and the gas mass in each cell is indicated by the color scale.
In the same panel, also shown by the grey line is the temperature evolution obtained by the one-zone model, where the gas density is assumed to increase at a fixed rate by using the free-fall time $t_\text{ff}$, i.e.,  $\rho / \dot{\rho} = t_\text{ff}$, with the same set of the thermal and chemical processes. 

In relatively metal-enriched cases with $Z/Z_\odot = 10^{-1}$ and $10^{-2}$,
the impact of CMB heating is remarkable both on the cloud temperature and on its morphology.
The cloud temperature is boosted to the CMB temperature floor.
At high redshifts $z\gtrsim 10$,
the gas evolves almost isothermally at the CMB temperature for $n\gtrsim 10^6~\mathrm{cm^{-3}}$, where the dust thermal emission efficiently dissipates the gas thermal energy, although with a large scatter in the temperature above the one-zone result at $n\lesssim 10^6~\mathrm{cm^{-3}}$,
due mainly to shock heating by colliding turbulent flows (Fig.~\ref{fig_rhoT_hist}).
The scatter becomes larger at lower metallicities since a shocked gas is harder to cool and remains longer at high temperatures.
The cloud morphology also changes due to the CMB heating at $Z/Z_\odot = 10^{-1}$ and $10^{-2}$ as seen in the density distribution in Fig.~\ref{fig_snapshot}.
Although the clouds are highly filamentary in shape at $z=0$ (bottom row), such structure becomes less prominent at $z=20$ (top row) as a consequence of the higher CMB temperature. 
At $z\lesssim10$, those filaments fragment and create small-scale structures, while CMB heating erases them at $z=20$. 
This indicates that the CMB with higher temperature increases the pressure support and the clouds become more stable against the gravitational collapse, forming a massive core near the center.

\begin{figure}
	\centering
		\includegraphics[width=7.5cm]{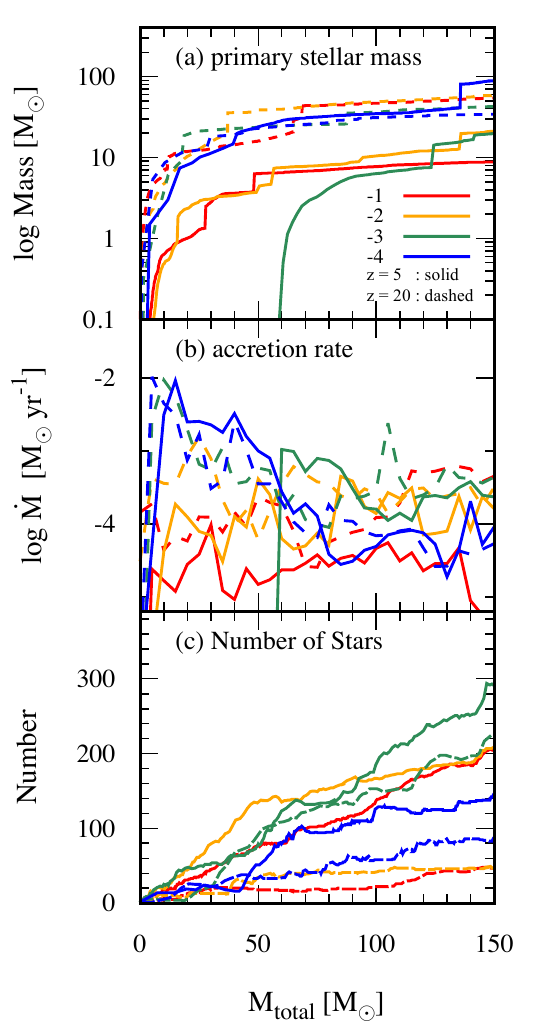}
		\caption{
		The evolution of (a) the stellar mass of the primary star, 
		(b) the mass accretion rate onto the primary star, and
		(c) the number of stars 
		as a function of the total stellar mass $M_\text{total}$.
		As $M_\text{total}$ monotonically increases with time,
		the horizontal axis can be regarded as a proxy for the time (Fig.~\ref{fig_mtot_evolution}).
		Different colors are for different metallicities, with
		$\log Z/Z_\odot = -1$ (red), $-2$ (orange), $-3$ (green), and $-4$ (blue lines),
        and the solid and dashed lines show the models with $z=5$ and $20$, respectively.
		}
		\label{fig_mass_evolutions}
\end{figure}

At lower metallicities $Z/Z_\odot = 10^{-3}$ and $10^{-4}$, although the CMB has weaker effects overall, at $z=20$ the cloud morphology is modified: substructures present at lower redshifts disappear and the cloud becomes more centrally concentrated (Fig.~\ref{fig_snapshot}).
As seen in Fig.~\ref{fig_rhoT_hist} (right two columns), the CMB affects the temperature evolution only at densities $n\lesssim 10^6~\mathrm{cm^{-3}}$, where the temperature decreases to $10~$K via HD cooling at $z=0$ and $5$ \citep{Nagakura&Omukai2005,Ripamonti2007,Hirano+2014}, while it remains at the CMB floor at $z=10$ and $20$.
This leads to differences in the mass of forming stars at high redshifts, as will be discussed in Section~\ref{sec::stellar_system} (see Fig.~\ref{fig_mass_evolutions}).

To see how the overall density structure of the collapsing cloud changes with the CMB temperature and metallicity, we plot in 
Fig.~\ref{fig_radial_profiles} the radial profiles of 
(a) the number density $n$, (b) temperature $T$, (c) mass infall rate $\dot{M}_\text{inf}$, and (d) enclosed mass $M_\text{enc}$ at the time of the first protostar formation as a function of the distance $r$ from the protostar at redshifts $z=5$ (solid) and $20$ (dashed).
We evaluate $\dot{M}_\text{inf}$ at the distance $r$ by
\begin{align}
\dot{M}_\text{inf} \equiv 4 \pi r^2 \rho v_\text{inf},
\end{align}
where $v_\text{inf}$ is the radial infall velocity of the gas.
The density profiles roughly follow the $n \propto r^{-2}$ law (panel a), consistent with the self-similar solution for the self-gravitating isothermal cloud collapse
\citep{Larson1969, Penston1969, OmukaiNishi1998}.
The density spike at $r\sim 10^3~$au when $Z/Z_\odot = 10^{-3}$ or $10^{-1}$ corresponds to formation of another dense core by fragmentation.
At the same metallicity, the temperature is higher at $z=20$ (dashed) than at $z=5$ (panel b) due to CMB heating.
As an example, when $Z/Z_\odot \gtrsim 10^{-2}$, the temperature at $r \gtrsim 100~$au is determined by the CMB floor of $\sim 60~$K at $z=20$.
Higher gas temperature at higher redshift enhances the mass infall rate (panel c) by about one order of magnitude from $z=5$ to $z=20$. 
This can be understood with the self-similar solution, where the mass infall rate is related to the gas temperature $T$ as
\citep{Whitworth&Summers1985,Foster+1993} 
\begin{align} \label{eq::Mdot}
\dot{M}_\text{inf} = 46.8 \frac{c_\text{s}^3}{G} = 6.4 \times 10^{-3}~M_\odot~\mathrm{yr^{-1}} 
\left ( \frac{T}{100~\mathrm{K}} \right )^{3/2},
\end{align}
indicating that higher mass infall rates are to be expected when the gas is at higher temperatures.
In the outer regions with $r \gtrsim 10^4~$au, in contrast, the accretion rates are $10^{-3}~M_\odot~\mathrm{yr^{-1}}$ for all metallicities, with smaller variations depending on the CMB temperature.
This reflects the fact that at large scales of $r \gtrsim 10^4$ -- $10^5~$au the cloud dynamics is controlled by
the overall gravity of the collapsing cloud, rather than by the thermal properties. 
In panel (d) the mass distribution appears to be more centrally concentrated in models with higher gas temperature, 
which is also consistent with the relation 
$M(r) \propto T r $
for the self-similar solution, whose density in the envelope is proportional to that of the singular isothermal sphere: $\rho \propto c_s^2/G r^2$.

\begin{figure*}
	\centering
		\includegraphics[width=15.5cm]{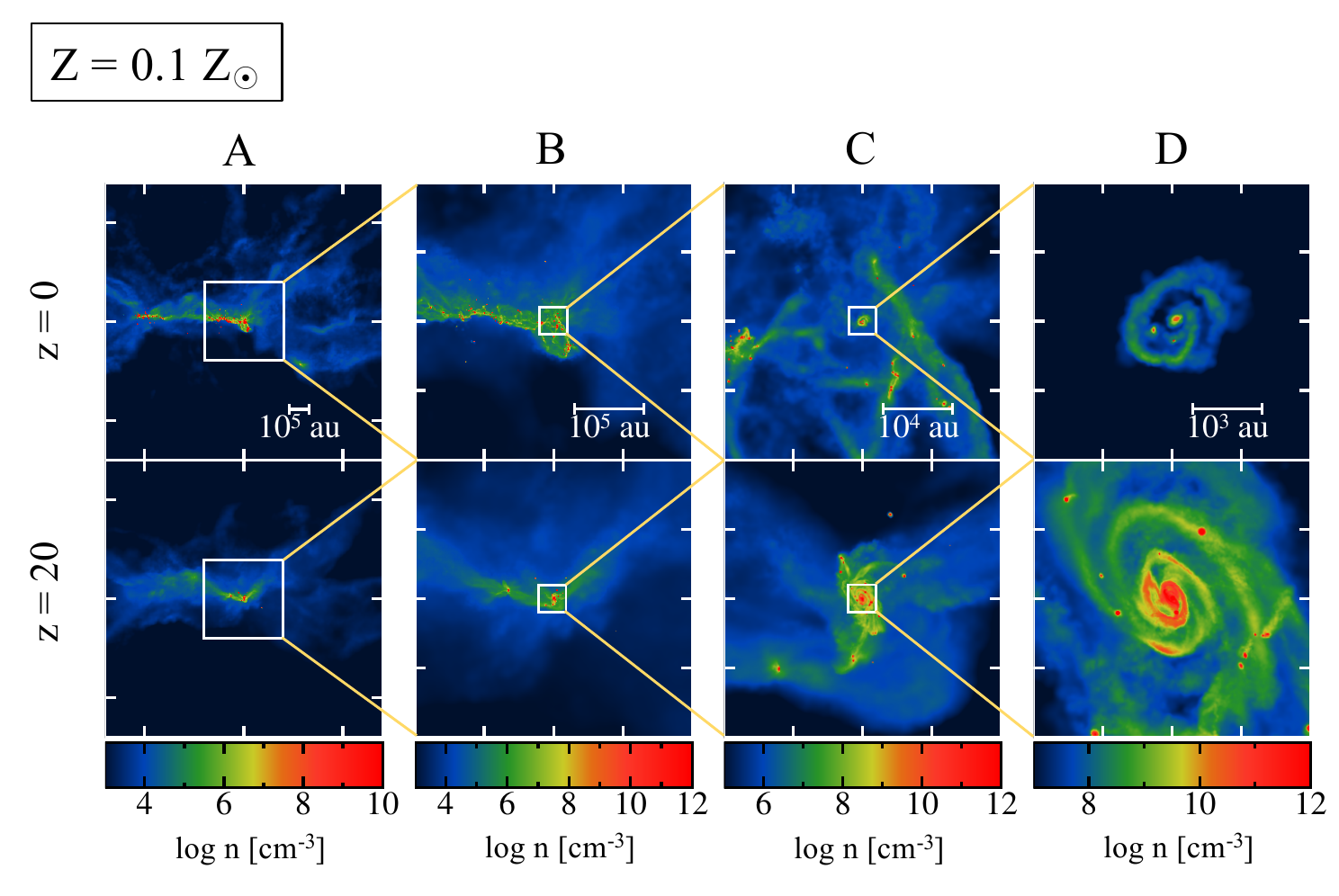}
		\caption{The density distributions at different spatial scales when the total stellar mass reaches $150~M_\odot$.
		The spatial scales becomes smaller from left to right.
		For comparison, top and bottom panels show the cases with different redshifts with $z=0$ (top) and $20$ (bottom) 
		but with the same metallicity $Z/Z_\odot = 0.1$.
		}
		\label{fig_snapshot_m150_zoomin}
\end{figure*}

\begin{figure*}
	\centering
		\includegraphics[width=15.5cm]{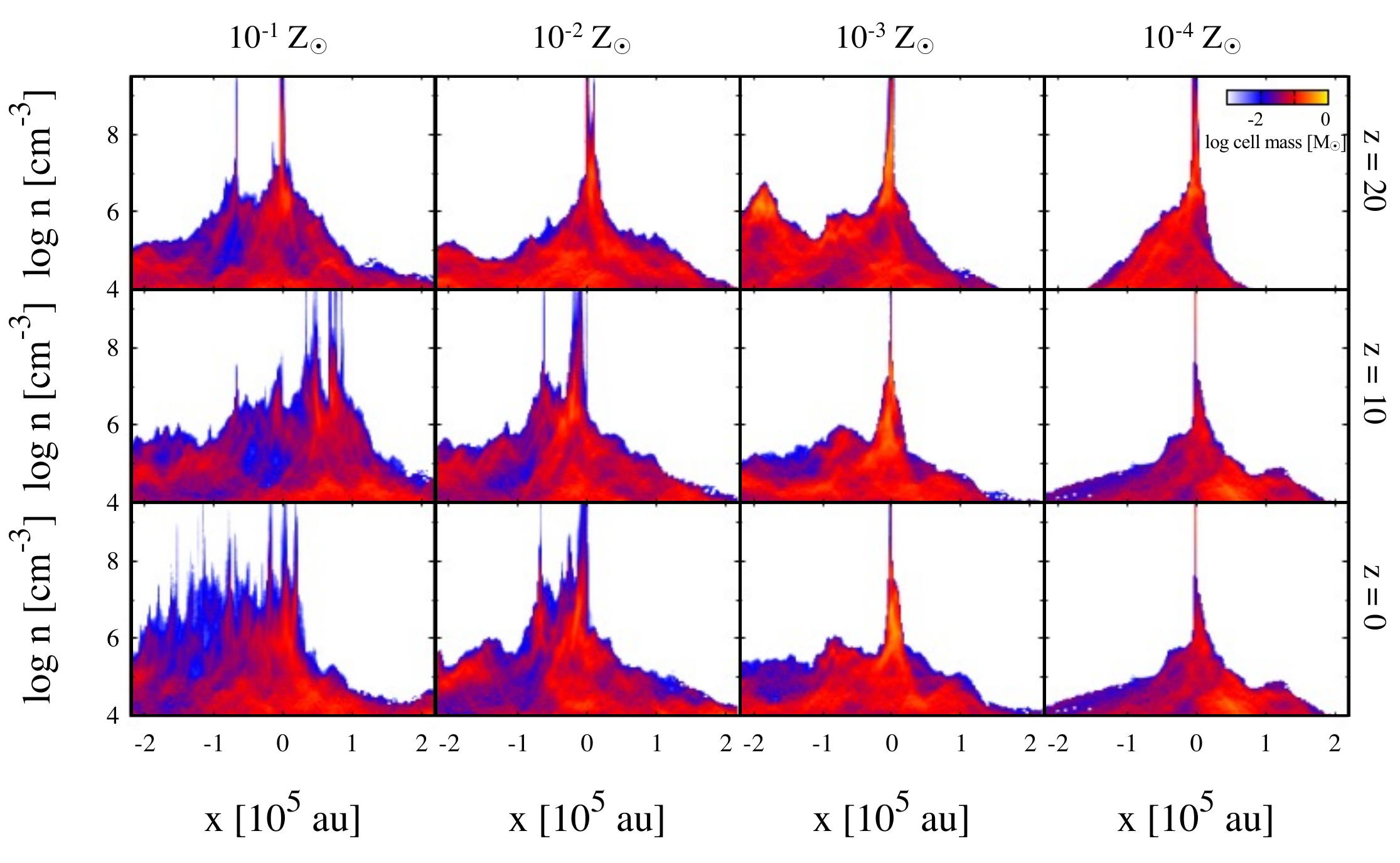}
		\caption{The position-density diagrams or "Clark's plot" 
		when the total stellar mass reaches $150~M_\odot$.
		In these plots, we project all the particles onto $x$-axis. 
		We divide the region by $200 \times 200$ cells and calculate the gas mass in each cell,
		which is indicated by the color.
		Emergence of fragments can be seen as spikes in this plot.
		}
		\label{fig_C_diagram}
\end{figure*}

\subsection{Accretion evolution of nascent stellar systems} \label{sec::stellar_system}
The clouds continue to fragment even after the protostar formation at the center and eventually yield multiple stars. 
To characterize the nature of the nascent stellar systems, we extend our calculation until the total stellar mass reaches $150~M_\odot$.
Around this epoch, massive stars with a few $10~M_\odot$ emerge and start emitting intense radiation after contraction to main-sequence stars \citep{Hosokawa+2009}.
Since radiation feedback from massive stars, not included in our simulation, would control the gas dynamics thereafter, 
we stop the calculation at this point.
Supposing that the radiation feedback suddenly kicks in and terminates the star cluster formation, we can regard the protostellar mass distribution at this moment as the stellar mass spectrum of the emerging star cluster since the mass distribution is almost fixed by then.
We, however, caution that the radiation feedback may not proceed in that way and can have significant impact on the resultant mass spectrum.
(see Section~\ref{sec::discussion}).

Fig.~\ref{fig_mtot_evolution} shows the growth of the total stellar mass, i.e., the sum of the mass of all stars, with time for the cases at $z=5$ (solid) and $20$ (dashed).
The origin of the time $t=0$ is taken at the epoch of the first protostar formation in the calculation. 
Stellar mass growth is more rapid at lower metallicity and higher CMB temperature, i.e., for higher gas temperature.

This agrees with the behavior of the mass infall rate shown in Fig.~\ref{fig_radial_profiles}(c), which is also higher for higher gas temperature. 
In relatively metal-enriched models with $Z/Z_\odot \gtrsim 10^{-2}$, 
the stellar mass increases more rapidly in the early phase of $M_* \lesssim 70~M_\odot$ at higher redshifts, 
while its rate in the later phase has a similar value among the cases with different redshifts.
This behavior can be understood from the mass infall rate shown in Fig.~\ref{fig_radial_profiles}(c):
at a high redshift, CMB heating raises the temperature and thus the infall rate in the inner region ($r\lesssim 10^4$--$10^5~$au), while the infall rate in the outer region has a similar value determined by the overall gravity of the cloud rather than by thermal pressure regardless of the redshift. 
In a later phase ($M_\text{tot}\gtrsim 100~M_\odot$), the increase of the total stellar mass is indeed similar between models with $z=5$ and $20$ (Fig.~\ref{fig_mtot_evolution}), reflecting the accretion of the gas from the outer region. 
In contrast to the early evolution described in Section~\ref{sec::collapse}, 
the CMB effect is more remarkable at low metallicities $Z/Z_\odot \lesssim 10^{-3}$.
For example, at $Z/Z_\odot=10^{-4}$ (blue lines), from the comparison between models at $z=5$ and $20$, 
the total star formation rate at $z=5$ suddenly decreases around $t=0.1t_\text{ff}$
and afterward becomes one order of magnitude smaller than at $z=20$. 
This is due to the temperature decrease from $\sim 100~$K to a few $10~$K by HD cooling at $z=5$,
which leads to a decrease of the mass growth rate by about one order of magnitude (see equation~\ref{eq::Mdot}).
Since the mass infall rate in the outer region is $10^{-3}~M_\odot~\mathrm{yr^{-1}}$,
the mass growth rate increases to this value again after a large enough mass accumulates at the cloud center as in the case with $Z/Z_\odot = 10^{-3}$.
Note that the stellar mass is still growing at the end of our simulation and further mass growth is expected for most models
except for that with $Z/Z_\odot = 10^{-4}$ and $z=5$. 
Since impact of the radiation feedback on the subsequent evolution and mass spectrum is difficult to predict, we stop the simulation at this stage
(see Section~\ref{sec::discussion}).

We next focus on CMB effects on the protostellar mass evolution and the number of protostars formed during star cluster formation.
Fig.~\ref{fig_mass_evolutions} shows
(a) the mass of the primary star, which is defined as the most massive star at the end of the calculation, i.e.,  when $M_\text{tot}=150~M_\odot$, 
(b) the mass accretion rate onto the primary star, and (c) the number of stars as a function of the total stellar mass $M_\text{tot}$.
Note that $M_\text{tot}$ monotonically increases with time and can be regarded as a proxy for the time for the growing stellar system.
Panel (a) shows that the growth of the primary star becomes more rapid at $z=20$ than at $z=5$.
This can also be seen in the behavior of the accretion rate (panel b), whose value 
at an early stage $M_\text{tot} \lesssim 50~M_\odot$ is one order of magnitude higher at $z=20$ than at $z=5$. 
Since the accreted material in such an early phase comes solely from the inside of the parental fragment, 
this suggests larger fragment mass-scale for higher CMB temperature owing to elevated Jeans mass at the fragmentation epoch.
When $Z/Z_\odot = 10^{-2}$ and $10^{-3}$, 
the accretion rates converge to $\sim 10^{-4}$ -- $10^{-3}~M_\odot~\mathrm{yr^{-1}}$, regardless of the redshift in the late phase of the evolution, when $M_\text{tot} \gtrsim 100~M_\odot$, consistent with the estimated mass infall rate shown in Fig.~\ref{fig_radial_profiles}(c). 
This indicates that the primary star at the center accretes the infalling gas exclusively without sharing it with other stars. 
In contrast, at the highest metallicity in our calculation $Z/Z_\odot=10^{-1}$, the mass accretion rate onto the primary star differs by one order of magnitude between the cases at $z=5$ and $20$ throughout the simulated period: it is typically $\sim 10^{-4}$--$10^{-3}~M_\odot~\mathrm{yr^{-1}}$ at $z=20$, while $\lesssim 10^{-5}~M_\odot~\mathrm{yr^{-1}}$ at $z=5$.
The latter is one order of magnitude smaller than the infall rate shown in Fig.~\ref{fig_radial_profiles}(c).
This discrepancy implies that the infalling material is shared among a number of massive stars formed by fragmentation.
On the other hand, at $z=20$, the high CMB floor suppresses fragmentation and allows the central massive star(s) to monopolize the infalling matter.
At the lowest metallicity of $Z/Z_\odot = 10^{-4}$, 
the mass evolution of the primary star hardly changes between the models with $z=5$ and $20$. 
This may sound contradicting with the fact that the growth of the total stellar mass becomes slower at $z=5$ due to HD cooling (Fig.~\ref{fig_mtot_evolution}).
When $z=20$, although the total accretion rate on the stars is higher, this material is shared among a larger number of massive stars at the cloud center.
In fact, we have found that the number of massive stars with $M_* \gtrsim 10~M_\odot$ is six at $z=20$, while it is only two at $z=5$.
Consequently, the accretion rate onto the primary star is similar in models with $z=5$ and $20$.

\begin{table}
\caption{Number of stars formed during the simulation.}
\begin{tabular*}{\columnwidth}{l@{\hspace*{30pt}}l@{\hspace*{30pt}}l@{\hspace*{30pt}}l@{\hspace*{30pt}}l}
\hline
$\log Z/Z_\odot$ & -1 & -2 & -3 & -4 \\[2pt]
\hline
\hline
$z=20$ & 47                      & 47                      & 225                     & 85                      \\[2pt]
$z=10$ & 111                     & 206                     & 245                     & 52                      \\[2pt]
$z=5$  & 204                     & 213                     & 291                     & 146                     \\[2pt]
$z=0$  & 288                     & 325                     & 266                     & 152                     \\[2pt]
\hline
\label{tab::nstar}
\end{tabular*}
\end{table}

Fig.~\ref{fig_mass_evolutions}(c) shows that CMB heating suppresses fragmentation, 
as the number of formed stars decreases with increasing CMB temperature at all metallicities. 
Table~\ref{tab::nstar} summarizes the number of protostars formed in our simulations.
In moderately metal-enriched cases $Z/Z_\odot \gtrsim 10^{-2}$, the number of stars at $z=20$ becomes 
by a factor of four smaller than that at $z=5$.
With lower metallicities $Z/Z_\odot \lesssim 10^{-3}$, although the CMB effect becomes less significant, 
the number of stars is reduced, while the growth of the primary star is not affected by CMB heating.
We summarize here the epoch when the CMB effect on star formation becomes significant depending on the metallicity. 
When $Z/Z_\odot \lesssim 10^{-3}$,
the results are very similar at $z=20$ and $z=10$ but are quite different when $z=5$, suggesting that there is a transition around $z=5$--$10$. 
In more metal-enriched models, when $Z/Z_\odot \gtrsim 10^{-2}$,
the number of stars gradually increases with decreasing redshift in the range of $0 < z < 20$.  
This indicates that CMB heating is still important at $z \simeq 5$ (see also Section~\ref{sec::spectrum}). 

\begin{figure*}
	\centering
		\includegraphics[width=18.cm]{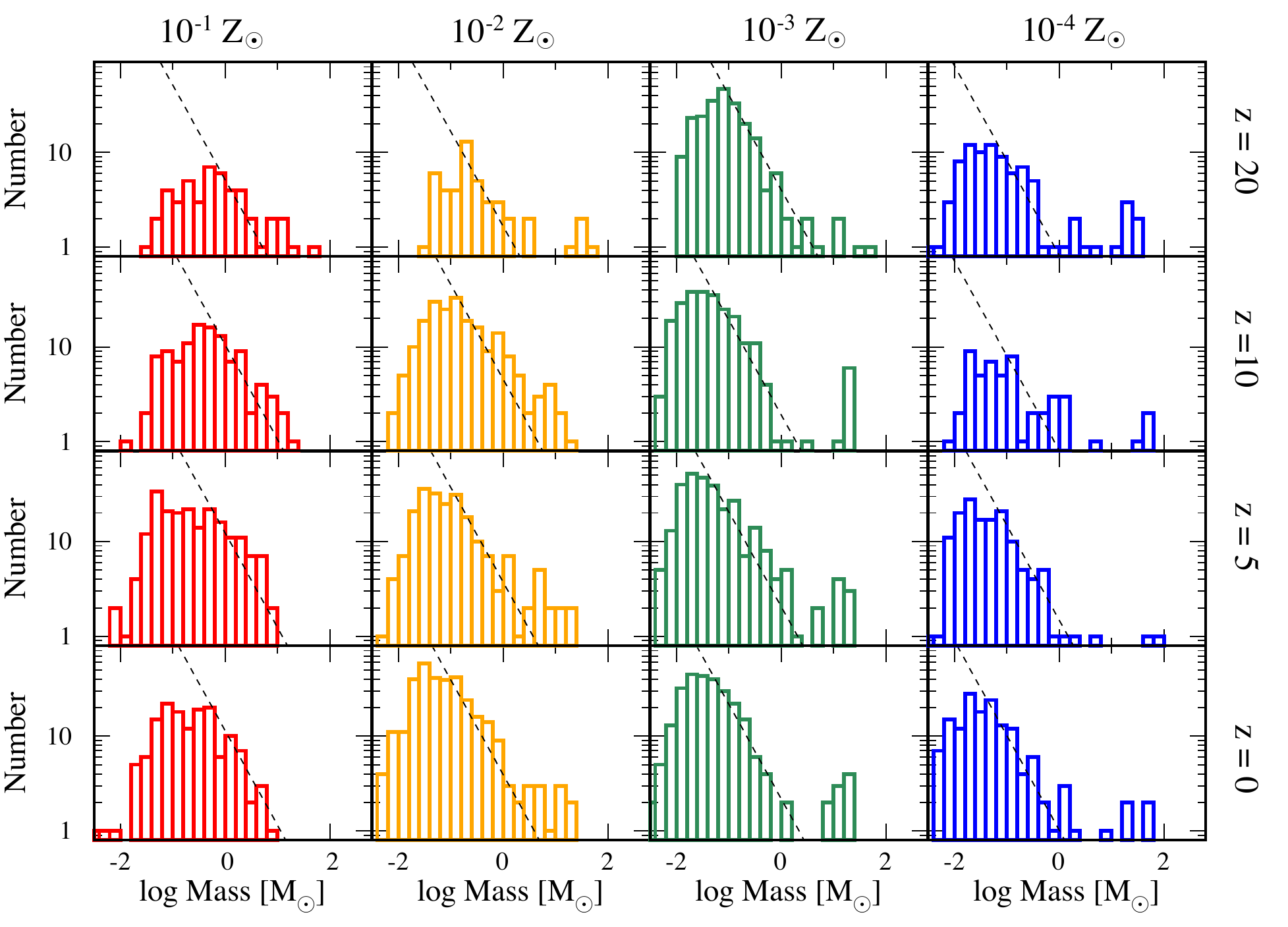}
		\caption{The stellar mass distributions when the total stellar mass reaches $150~M_\odot$ for the cases with various metallicities (columns) and redshifts (rows).
		The black dashed line represents the power-law with the exponent of $-1$ 
		(equation~\ref{eq:mass_spectrum}).}
		\label{fig_mass_spectrum}
\end{figure*}
\begin{figure*}
	\centering
		\includegraphics[width=16.cm]{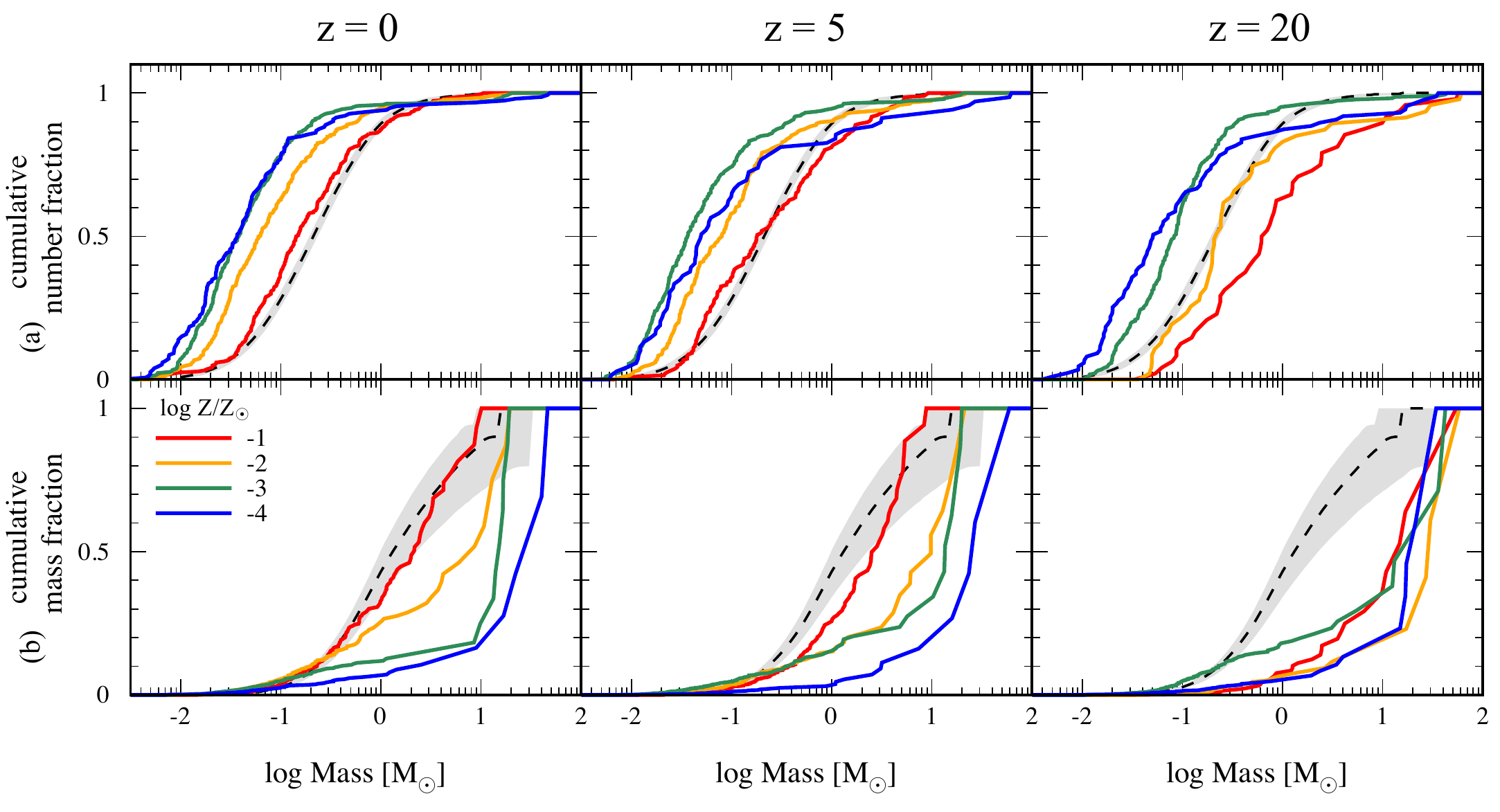}
		\caption{The cumulative fractions of the number of stars (top) and stellar mass (bottom) for different metallicities and redshifts.
		We integrate the number of stars and mass from the low-mass end and normalize them, respectively, by the total stellar number and mass.
		The black dashed lines show the cumulative fractions for the Chabrier IMF with the maximum stellar mass $100~M_\odot$ \citep{Chabrier2003},
		where we stocastically sample the mass spectrum assuming the total stellar mass is $150~M_\odot$.
        The shaded region represents $1\sigma$ variance, 
        coming from the fact that we do not fully sample the IMF at the massive end.
		Colors indicate different metallicities	for $\log Z/Z_\odot = -1$ (red), $-2$ (orange), $-3$ (green), and $-4$ (blue).
		Different column shows the cumulative fraction at different redshifts.}
		\label{fig_cumulative_mass_spectrum}
\end{figure*}

CMB heating also affects the spatial distribution of protostars at high metallicities, when $Z/Z_\odot \gtrsim 10^{-2}$.
The positions of the protostars at the final snapshot of $M_{\rm total}=150~M_\odot$ are overplotted in Fig.~\ref{fig_snapshot}, where asterisks (points) represent protostars whose masses are larger (smaller, respectively) than $1~M_\odot$.
Note that the gas distribution on these scales hardly changes since the first protostar formation until the end of the simulation. 
When $Z/Z_\odot = 10^{-1}$, the spatial distribution of the protostars is affected by CMB heating:
at $z=20$ cloud fragmentation is suppressed by the higher CMB temperature and protostars are formed only around the cloud center, while at $z=0$ a filamentary structure develops and protostars formed by filament fragmentation are distributed along the filaments over the entire initial cloud core scale of $\sim$pc.
At $Z/Z_\odot = 10^{-2}$, without a clear filamentary structure, the cloud hardly fragments. 
As a result, only a single massive core appears and massive stars are formed around the center regardless of the CMB heating.
When $Z/Z_\odot = 10^{-3}$, the cloud fragments into two massive cores, one of which is stabilized by CMB heating at $z=20$. In this case, stars form only in the other core.
At lower redshifts at $z \lesssim 10$, stars form in both cores due to weaker CMB heating.
When $Z/Z_\odot \lesssim 10^{-3}$, most of the massive stars are located around the cloud center while some low-mass stars are ejected from the system as a result of multi-body interaction.
In these cases, a massive and compact gas disk forms around the central protostellar system. 
Inside the disk, dust cooling is effective (Fig.~\ref{fig_rhoT_hist}) and induces vigorous fragmentation, yielding a number of low-mass stars \citep[e.g.][]{Tanaka&Omukai2014}.
Close encounters among the stars causes ejection of low-mass stars, resulting in their spatially extended distribution. 
In such low-metallicity cases,
fragmentation of the circumstellar disks is not affected by CMB heating as
the temperature in the disk is higher than the CMB temperature.  
Therefore, stellar ejection is observed regardless of the redshift.
The ejection of low-mass stars is particularly significant when $Z/Z_\odot =10^{-4}$.
The spatial distribution of low-mass stars shows no clear correlation with the CMB temperature owing to stochastic nature of the ejection process. 

We have seen that when $Z/Z_\odot=10^{-1}$, 
CMB heating significantly alters the mass and the spatial distribution of the protostars.
To see its effects on the growth of the protostars, we plot in Fig.~\ref{fig_snapshot_m150_zoomin} the corresponding density structure around the protostars at various scales for $z=0$ (top) and $20$ (bottom).
At large scales of $\gtrsim 10^5~$au (column A), the overall density structure is similar between $z=0$ and $20$:
collision of large-scale turbulent flows creates a filamentary structure with $\lesssim 10^4~\mathrm{cm^{-3}}$.
Fine structures inside the filament, however, are significantly different between the two cases, as shown in column B:
the filament fragments when $z=0$, while it does not
at $z=20$ due to higher temperature and thus pressure opposing the gravitational collapse.

\begin{figure}
	\centering
		\includegraphics[width=8.5cm]{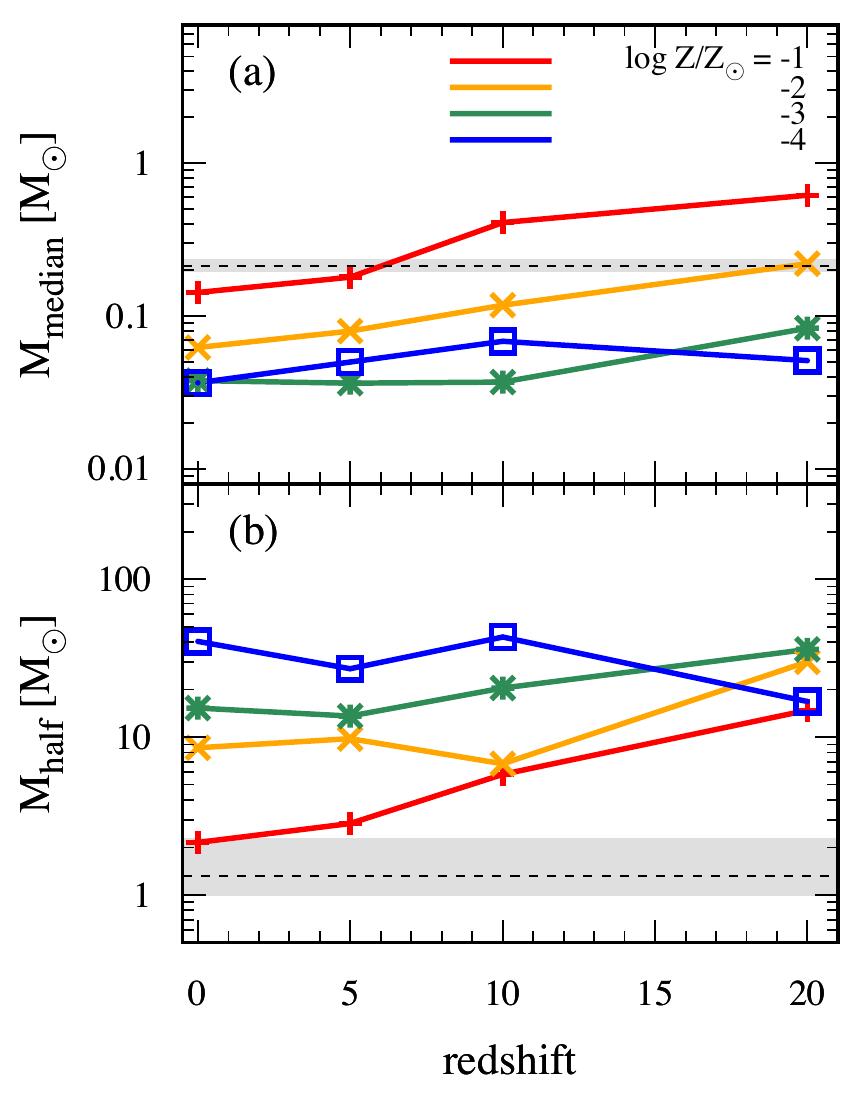}
		\caption{(a) The median mass $M_\text{median}$ and (b) $M_\text{half}$ 
		as a function of redshift for different metallicities $\log Z/Z_\odot=-1$ (red), $-2$ (orange), $-3$ (green), and $-4$ (blue).
		We define $M_\text{half}$ as the mass 
		below or above which half the total stellar mass is contained.
		The dashed lines represent $M_\text{median}$ and $M_\text{half}$ 
		for the Chabrier IMF with the maximum mass of $100~M_\odot$
		and the shaded regions indicate 1-$\sigma$ variance.
		}
		\label{fig_half_mass_scales}
\end{figure}
Suppression of filament fragmentation also changes the density structure at smaller scales, $10^3$--$10^4~$au. 
In column C, 
at $z=0$ (top panel) the most massive star and the surrounding material are located outside the filament. 
This clearly shows how fragmentation hinders the growth of the primary star by quenching the mass supply.  
In contrast, at $z=20$ (bottom panel), the circumstellar gas is directly connected to the filamentary structure and the continuous gas supply along it allows the protostar to grow efficiently.
The infalling gas accumulates on the circumstellar disk, increasing the size and mass of the disk.
Column D shows that the disk is  $10^3~$au in size at $z=0$, while it is one order of magnitude larger at $z=20$. 
In the former case, the disk is gravitationally stable and 
low-mass stars are mainly formed via filament fragmentation on larger scales, while in the latter case, the disk is gravitationally unstable and fragments, yielding a number of low-mass stars close to the central region (see also Fig.~\ref{fig_snapshot}).
At lower metallicities, $Z/Z_\odot \lesssim 10^{-2}$,
the large-scale filament tends to be stable due to inefficient cooling and the primary star grows in a similar way as 
in the $10^{-1}~Z_\odot$ model at $z=20$ discussed above.

Fig.~\ref{fig_C_diagram} shows the position-density diagram, or the so-called Clark's plot \citep[after Fig.3 of][]{Clark+2008}, at the final snapshot $M_{\rm total}=150~M_\odot$ for various metallicities and redshifts.
Here the particle distribution is projected onto the $x$-axis and the $n$-$x$ plane is divided by $200 \times 200$ cells.
The color represents the gas mass in each cell.
Emergence of fragments can be seen as spikes in this plot.  
In the highest metallicity model ($Z/Z_\odot = 0.1$), sharp spikes are visible at $z=0$. 
This indicates that a filament is fragmenting at the inter-spike density of $\sim 10^7~\mathrm{cm^{-3}}$. 
Above this density, $\gamma$ exceeds unity due to inefficient cooling (Fig.~\ref{fig_rhoT_hist}), and the development of filaments and their fragmentation are suppressed \citep[e.g.][]{Tsuribe&Omukai2006}.
At this point, the fragmentation mass scale can be estimated as \citep{Schneider&Omukai2010},
\begin{align} \label{eq::MJ}
M_\text{frag} \sim 0.2~M_\odot 
\left ( \frac{T_\text{gas}}{10~\mathrm{K}} \right )^{1.5} 
\left ( \frac{n}{10^7~\mathrm{cm^{-3}}} \right )^{-0.5},
\end{align}
corresponding to the typical protostellar mass
(see Section~\ref{sec::spectrum}).
The number of spikes decreases with increasing redshift.
At $z=20$, the filaments have a typical density $10^5$--$10^6~\mathrm{cm^{-3}}$ and a temperature $60$--$100$~K
(see Figs.~\ref{fig_snapshot_Tgas} and \ref{fig_rhoT_hist}).
Owing to low density and high temperature inside the filament, the fragmentation mass scale becomes as high as $20~M_\odot$, indicating that low-mass stars with $M_* \lesssim 1~M_\odot$ are not formed by filament fragmentation.
The situation is similar in low-metallicity models with $Z/Z_\odot \lesssim 10^{-2}$, 
where only a few density peaks, i.e., fragments, appear around the cloud center, 
surrounded by a low-density envelope with $n \sim 10^5~\mathrm{cm^{-3}}$.
The fragmentation scale is a few $\times$ 10 $M_{\odot}$, indicating that massive stars will be formed inside the fragments (see Section~\ref{sec::spectrum}).

\subsection{Mass spectrum} \label{sec::spectrum}
The mass distribution of stars at the end of the calculation is shown in Fig.~\ref{fig_mass_spectrum}.
As discussed in paper I, the stellar mass spectrum consists of two components at low metallicities.
The low-mass component has the universal power-law shape (black dashed)
\begin{align} \label{eq:mass_spectrum}
\frac{\mathrm{d}N}{\mathrm{d}\log M_*} \propto M_*^{-1},
\end{align}
at the high-mass end extending to $\lesssim 10~M_{\odot}$ and peaks at $0.01$--$0.1~M_\odot$ in the logarithmic mass bin.  
In addition to this, a massive stellar component with roughly log-flat spectrum appears in the range $M_* \sim 10$--$100~M_\odot$ when the metallicity is very low $Z/Z_\odot \lesssim 10^{-2}$.
Those massive stars preferentially grow in mass anchoring at the center of the collapsing cloud.
In the case with moderately high metallicity ($Z/Z_\odot=10^{-1}$) and low redshift ($z=0$ or $5$), such stars do not appear due to vigorous fragmentation of the cloud and the stellar mass spectrum can be described by the single component of equation~\eqref{eq:mass_spectrum}. 

To compare the mass spectra in a more quantitative way, 
we plot the cumulative distributions of (a) stellar number and (b) mass
in Fig.~\ref{fig_cumulative_mass_spectrum}, 
where those quantities are summed up from the lower-mass end and normalized by the total stellar number and mass, respectively.
The black dashed lines show the predictions for the Chabrier IMF with the maximum stellar mass of $100~M_\odot$,
where we stochastically sample the mass spectrum following the Chabrier IMF 
assuming the total stellar mass is $150~M_\odot$.
Since the total stellar mass is small in our final snapshot, 
the IMF cannot be fully sampled especially at the high mass end.
The shaded regions represent the $1\sigma$ variance of the cumulative fractions,
where we randomly generate $10^5$ realizations of the mass spectrum.
We can observe that the cumulative number fraction has negligible variance 
since the number fraction in massive stars is very small for the Chabrier IMF.
On the other hand, larger variance appears at the high mass end in the cumulative mass fraction,
since massive stars account for a substantial mass fraction,
e.g., the mass fraction of those with $M_* > 10~M_\odot$ amounting to 20\%.
Thus, the stochasticity of the number of massive stars introduces larger error in the mass fraction.

At very low metallicities $Z/Z_\odot \lesssim 10^{-3}$ (blue and green lines), 
the CMB does not affect the mass spectrum significantly.
While low-mass stars are more numerous than in the Chabrier IMF (top panel), they occupy only a small fraction in terms of mass (bottom panel) regardless of the CMB temperature.
At higher metallicities $Z/Z_\odot \gtrsim 10^{-2}$ (yellow and red lines), the CMB impact on the mass spectrum is more prominent.
CMB heating suppresses the formation of low-mass stars with $0.01 < M_*/M_\odot < 1$ at $z=20$
and more than 50\% (80\%) of the total mass is in massive objects with $M_* > 10~M_\odot$ for $Z/Z_\odot = 10^{-1}$ ($10^{-2}$, respectively).
Meanwhile, at $z=5$, the mass fraction in massive objects becomes 50\% for $Z/Z_\odot = 10^{-2}$ and
no stars with $M_* > 10~M_\odot$ appear in $Z/Z_\odot=10^{-1}$.
This demonstrates that CMB heating makes the mass function significantly more top-heavy than the Chabrier IMF, similar to low values of metallicity with $Z/Z_\odot \lesssim 10^{-3}$.
When the metallicity reaches $Z/Z_\odot = 10^{-1}$ and the redshift becomes $z = 0$, 
the mass function finally follows a Chabrier-(or Salpeter-)like IMF
both in terms of the number and mass fractions. 

Although the mass spectrum is top-heavy both in low-metallicity and in high-redshift environments, the number of low-mass stars is much smaller at high redshift: low-mass stars with $M_* < 0.1~M_\odot$ are less than 20\% in number at $z=20$ but more than 80\% when $Z/Z_\odot \lesssim 10^{-3}$ at all the redshift considered here (Fig.~\ref{fig_cumulative_mass_spectrum} a).
This difference comes from the difference in the thermal evolution shown in Fig.~\ref{fig_rhoT_hist}.
At high redshifts ($z=20$ and $Z/Z_\odot \gtrsim 10^{-2}$), the gas is almost isothermal at the CMB temperature. 
On the other hand, in low-metallicity cases ($Z/Z_\odot \lesssim 10^{-3}$ at any redshift)
the temperature first gradually increases and then suddenly drops at $n \sim 10^{8}~\mathrm{cm^{-3}}$ by dust cooling. 
This temperature drop promotes vigorous fragmentation of the disks and yield a number of low-mass stars \citep{Tanaka&Omukai2014}.
This is also consistent with the numerical experiments by 
\citet{Li+2003} who investigated cloud fragmentation with polytropic equations of state and showed that fragmentation is less frequent in the isothermal case ($\gamma=1$) than in the cases with temperature decreasing with density ($\gamma < 1$). 

In Fig.~\ref{fig_half_mass_scales}, we show the redshift evolution of two typical stellar mass scales: (a) the median mass $M_\text{median}$ and (b) the half mass scale $M_\text{half}$.
Here the latter is defined so as the total mass in stars more massive than $M_\text{half}$ equals half the total mass.
Those values for the Chabrier IMF are also indicated by 
the dashed lines. 
Note that $M_\text{median}$ ($M_\text{half}$, respectively) is sensitive to the number of low-mass (massive) stars. 
CMB heating has the strongest impact in the highest metallicity model $Z/Z_\odot = 10^{-1}$. 
In this and also $Z/Z_\odot = 10^{-2}$ models, 
both $M_\text{median}$ and $M_\text{half}$ increase with redshift, reflecting more massive fragmentation scales ($M_\text{median}$) and more efficient growth by accretion for massive stars ($M_\text{half}$) at higher redshifts.
At lower metallicities ($Z/Z_\odot = 10^{-3}$ and $10^{-4}$), the CMB has negligible impact both on $M_\text{median}$ and $M_\text{half}$ in the plotted redshift range. 
Rather, stochastic processes are more important in determining stellar masses than the CMB effect in those cases, where low-mass stars are mostly formed via disk fragmentation in crowded central regions and
are easily ejected from the birth sites by dynamical interaction with other stars 
(Fig.~\ref{fig_snapshot}, right two columns in the upper panel).
The number of low-mass stars are thus vulnerable to stochasticity and the median mass has no systematic dependence on redshift.
The growth of massive stars is also stochastic.
The total mass at the upper end is dominated by a small number of massive stars ($M_* \gtrsim 10~M_\odot$).
Since massive stars tend to be in triple or multiple systems, dynamical interaction among them affects chaotically their accretion growth.
This may cause variations of order of a few tens of $M_\odot$ in $M_\text{half}$.
The redshift variation of $M_\text{half}$ is within this variance in the cases $Z/Z_\odot = 10^{-3}$ and $10^{-4}$ and no clear dependence on the redshift can be observed.

\begin{figure}
	\centering
		\includegraphics[width=8.5cm]{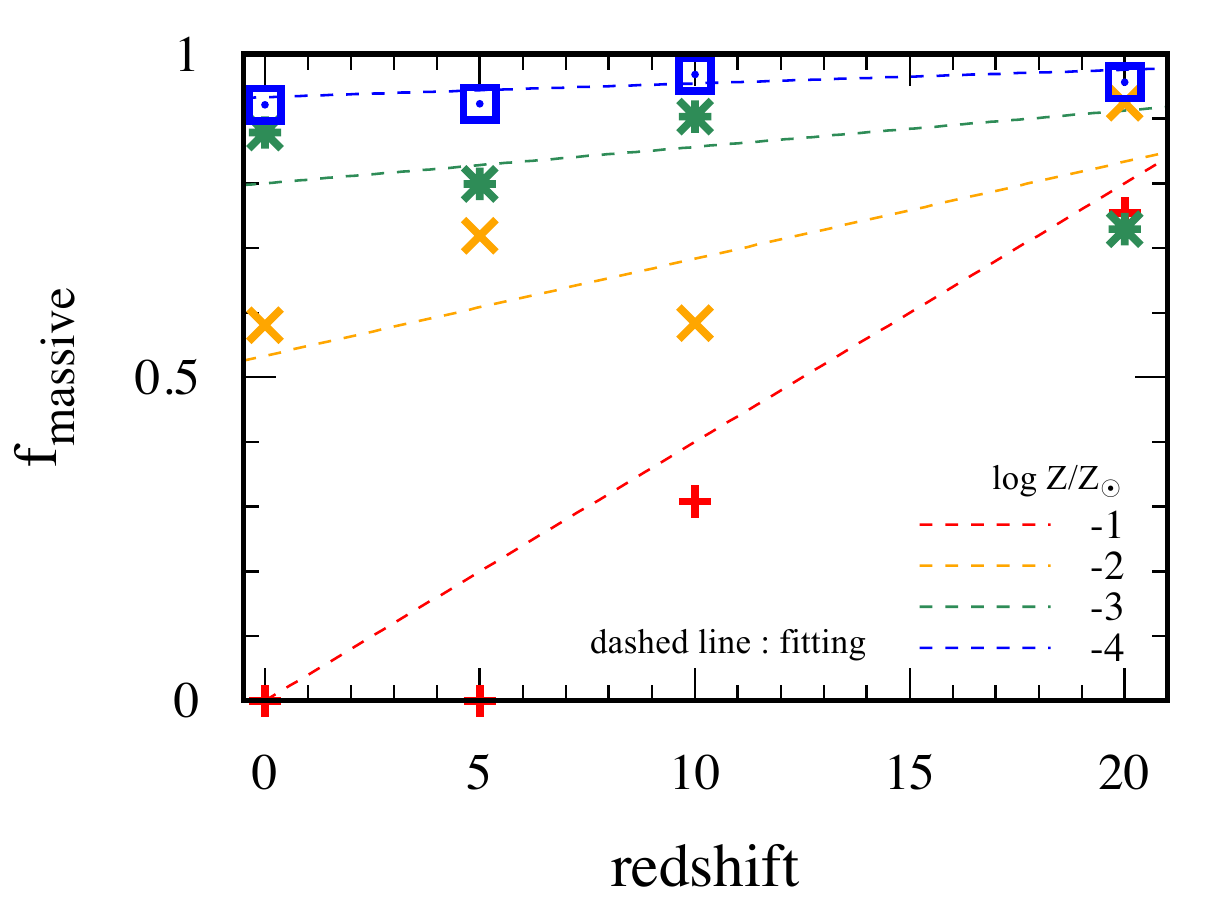}
		\caption{
		The mass fraction in the log-flat massive stellar component $f_\text{massive}$ as a function of the redshift when the total stellar mass reaches $150~M_\odot$ for metallicities $\log Z/Z_\odot = -1$ (red), $-2$ (orange), $-3$ (green), and $-4$ (blue).
		The dashed lines show the fitting expression (equation~\ref{eq::fitting}) for the redshift evolution of the mass in the massive stellar component.
		}
		\label{fig_massive_component}
\end{figure}

We here examine how the mass distribution deviates from the present-day IMF.
As shown in Fig.~\ref{fig_mass_spectrum},
the mass spectrum is composed of a Salpeter-like part at $M_*\lesssim 5$--$10~M_\odot$ and
a massive component with a log-flat mass distribution with $10~M_\odot \lesssim M_* \lesssim 100~M_\odot$.
We quantify the mass fraction in the massive component by fitting the Salpeter-like component 
with the power law $\propto M_*^{-1}$ in the range 
$M_\text{median} < M_* < 5~M_\odot$, as shown by the black dashed lines in Fig.~\ref{fig_mass_spectrum}, and then regarding  
stars more massive than the mass above which 
the power-law distribution gives less than unity to belong to the massive component.
Fig.~\ref{fig_massive_component} shows the mass fraction in the massive component as a function of the redshift.
The different symbols indicate the different metallicities $\log Z/Z_\odot = -1$ (red), $-2$ (orange), $-3$ (green), and $-4$ (blue).
As expected, the total mass in the massive component increases with increasing redshift or decreasing metallicity, i.e., higher gas temperature.
This result can be roughly fitted with the dashed lines given by:
\begin{align} \label{eq::fitting}
f_\text{massive} &= 1.07 * \left(1 - 2^{x} \right) + 0.04 \times 2.67^ {x} \times z, \\
{\rm with}~~
x &= 1 + \log Z/Z_\odot, \nonumber
\end{align}
where $z$ is the redshift under consideration. 
In the limit $x \rightarrow -\infty$ (i.e., $Z=0$), 
this expression successfully reproduces the mass spectrum of the first stars obtained by previous studies, where the spectrum is composed purely of the massive log-flat component \citep[e.g.][]{Susa2019, Chon+2021b}.
Note also that in this expression the massive component disappears when $\log Z/Z_\odot \rightarrow -1$ and $z \rightarrow 0$ as expected.

Finally, we remark the different spatial distributions of stars in the Salpeter-like and massive log-flat components.
Massive stars tend to be found around the cloud center and constitute a binary or higher-order multiple system,
as it is frequently found in simulations for primordial star-forming clouds 
\citep{Stacy+2016, Chon+2018, Chon+2019, Susa2019, Sugimura+2020, Matsukoba+2021}.
Around the binary/multiple system, 
there is a massive gas disk, which feeds the massive member stars with gas \citep[e.g.][]{Chon&Omukai2020}, leading to the top-heavy log-flat mass distribution.
On the other hand, 
being formed via dust-cooling induced fragmentation, the fate of the low-mass Salpeter-like component
depends on the fragmentation mode:
when they are generated by disk fragmentation ($z\gtrsim10$ or $Z/Z_\odot \lesssim 10^{-2}$), 
most of them are quickly ejected from the central region and have spatially extended distribution.
In the case of filament fragmentation ($Z/Z_\odot = 0.1$ and $z \lesssim 5$), they are more spatially concentrated and located along the filament
(see Fig.~\ref{fig_snapshot}).

\section{Discussion} \label{sec::discussion}
We have simulated star cluster formation for a wide range of the metallicities and CMB temperatures.
Dust cooling induces vigorous fragmentation of circumstellar disks, yielding a large number of low-mass stars
\citep{Tsuribe&Omukai2006, Clark+2008, Jappsen+2009, Dopcke+2013, Chiaki+2016, Chiaki+2021, Shima&Hosokawa2021}.
In cases with negligible CMB effect ($z=0$), our results are broadly consistent with previous studies on metallicity effects on the mass spectrum.
While dust cooling enables low-mass star formation, a dominant fraction of mass is still in massive stars when $Z/Z_\odot \lesssim 10^{-2}$.
The mass function does not become a present-day Salpeter-like IMF 
until the metallicity is as high as $Z/Z_\odot \sim 10^{-2}$ -- $10^{-1}$, consistent with the result in Paper~I.

The CMB effect is more prominent in higher metallicity cases, where cooling is more efficient and the temperature hits the CMB floor.
Higher CMB temperature significantly modifies the cloud structure on scales of $0.1$--$1~$pc. 
Development of the filamentary structure and its fragmentation are suppressed (as seen in models at $z=20$ with $Z/Z_\odot = 10^{-1}$), and the fragment number is reduced. 
Our results are consistent with those obtained by \citet{Smith+2009},
who performed hydrodynamical simulations starting from cosmological initial condition and followed the cloud evolution in a metallicity range $Z/Z_\odot \lesssim 10^{-2}$.
They found that the cloud morphology is significantly affected by the CMB temperature floor in mildly metal-enriched models with $Z/Z_\odot \sim 10^{-2}$.
CMB heating strongly suppresses the development of a filamentary structure and its fragmentation, which is observed when the effects of CMB heating are not considered.
While their calculation is terminated when the density reaches $10^{11}~\mathrm{cm^{-3}}$, we have followed the evolution until a far more advanced phase 
when protostars form inside the fragments and the total stellar mass reaches $150~M_\odot$. 
This enables us to see how the CMB heating and the suppression of the cloud fragmentation affect the stellar mass spectrum leading to a more top-heavy distribution.

Our results also qualitatively agree with the semi-analytical calculation by \citet{Schneider&Omukai2010}, which claimed that warm CMB suppresses cloud fragmentation at a high redshift and increases the fragmentation mass scale. 
The mass scale of filament fragmentation in our calculation indeed strongly depends on the CMB temperature.
Although the fragmentation mass scale at $z=0$ in \citet{Schneider&Omukai2010} is similar to the median mass scale in this study, it becomes one order of magnitude larger than in our calculation at $z \gtrsim 5$. 
This is because low-mass stars are still produced owing to the fragmentation of circumstellar disks while filament fragmentation is suppressed by the CMB heating.
For example, even at redshift as high as $z=20$, 
stars are distributed in a broad mass range when $Z/Z_\odot = 10^{-2}$ or $10^{-1}$ (Fig. \ref{fig_mass_spectrum}).
This effect, which is not considered in \citet{Schneider&Omukai2010}, makes the typical stellar mass smaller by one order of magnitude \citep[e.g.][]{Riaz+2020}.

Our result indicates that the critical metallicity for the IMF transition ($Z_\text{crit}$) increases with the increasing redshift due to the higher CMB heating rate at higher redshift. This makes $Z_\text{crit}/Z_\odot\sim 10^{-2}$ at $z=0$ and $\gtrsim 10^{-1}$ at $z \gtrsim 5$ (see Fig.~\ref{fig_cumulative_mass_spectrum}). In a similar analysis, \citet{Bromm+2001b} have also derived $Z_\text{crit}$, which is far smaller than ours and has an opposite trend with redshift. This stems from the difference in physical conditions of the clouds considered in two studies.  While ours is molecular-cooling clouds without any irradiation other than by the CMB, they considered H$_2$-photodissociated clouds without the CMB effect. At higher redshift, the higher temperature floor by the CMB nullifies cooling and requires more metals to have any effect in our case.
On the other hand, in their analysis, higher density at virialization at higher redshift results in higher cooling rate by metal fine-structure lines, i.e., a smaller amount of metals can make a difference.  

Our calculation is terminated when the total stellar mass reaches $150~M_\odot$.
Around this epoch, the most massive stars reach a mass of a few $10~M_\odot$ and, after some interval for the Kelvin-Helmholtz contraction, start emitting copious amount of ionizing photons \citep{Hosokawa+2009}. 
The surrounding gas is heated up by radiation, eventually quenching further mass supply \citep[][]{Peters+2010, Dale+2012, Walch+2012, Geen+2018, He+2019, Fukushima+2020a}.
Hence, this subsequent evolutionary phase needs to be followed, 
accounting for ionizing radiation feedback, which is expected to change the thermal properties and dynamics of accretion flows, and thereby to modify the stellar mass spectrum. 
\citet{He+2019} have carried out radiation hydrodynamics simulation of star cluster formation and found that the mass accretion rate onto the forming stars decreases
by stellar radiation feedback although the shape of the mass spectrum, in particular, its slope at the high-mass end, is not significantly affected.
They have also provided a fitting function for the star formation efficiency as a function of the initial mass $M_\text{cloud}$ and density $n_\text{init}$ of the cloud,
\begin{align}
f_* = 0.032 \left ( \frac{M_\text{cloud}}{6300~M_\odot} \right )^{0.38}
\left ( 1 + \frac{n_\text{init}}{10^3~\mathrm{cm^{-3}}} \right )^{0.91}.
\end{align}
Inserting $M_\text{cloud} = 6300~M_\odot$ and $n_\text{init}=10^3~\mathrm{cm^{-3}}$ adopted in our simulation, we can expect that the star formation is terminated when the total stellar mass approaches $\sim 200~M_\odot$, comparable to the total stellar mass at the end of our calculation.
Therefore, we expect that the mass function obtained in our simulation, which is terminated at $M_\text{tot} = 150~M_\odot$,
can be quantitatively similar to the IMF that is realized after the radiation feedback quenches the accretion flows.

Even in protostellar phases, 
stellar radiation feedback could modify the mass distribution at the low-mass end by suppressing fragmentation at small scales.
In the case of present-day star formation, 
this feedback is considered to play a role in reproducing the observed small number of brown dwarfs by quenching the formation of objects smaller than $0.01~M_\odot$
\citep[e.g.][]{Bate2009, Myers+2011, Bate2012, Krumholz+2012, Bate2019}.
The impact of protostellar radiation feedback has also been studied for 
low-metallicity cases by several authors. 
\citet{Omukai+2010} estimated the extent of the radiation heating effect in the outer envelope of a star-forming cloud core by means of spherically symmetric hydrodynamics calculations and concluded that it has negligible impact on fragmentation when $Z/Z_\odot \lesssim 10^{-2}$.
By performing three-dimensional simulations,
\citet{Safranek-Shrader+2016} demonstrated that thermal feedback significantly increases the gas and dust temperature in circumstellar disks, thereby suppressing disk fragmentation at $Z/Z_\odot = 10^{-2}$.
Here, we estimate the impact of protostellar radiation on the surrounding environment 
at high redshift with high CMB temperature,
where heating by protostellar accretion luminosity is dominant only in the close vicinity of the stars.
The heating rate of dust grains by stellar radiation is $\kappa_\text{gr} L_\text{acc}/4 \pi r^2$, 
where the accretion luminosity is;
\begin{align}
L_\text{acc} &= f_\text{acc} \frac{GM_*\dot{M}_\text{acc}}{R_*} \nonumber \\
&= 5.4 \times 10^{37}~\mathrm{erg~s^{-1}}  \nonumber \\
& \;\;\; \left (\frac{M_*}{10~M_\odot} \right )
\left (\frac{\dot{M}_\text{acc}}{3\times10^{-4}~M_\odot~\mathrm{yr^{-1}}} \right ) \left (\frac{R_*}{5~R_\odot} \right )^{-1},
\end{align}
and $M_*$ is the stellar mass, $\dot{M}_\text{acc}$ is the mass accretion rate onto the star,
$R_*$ is the stellar radius,
and $f_\text{acc}$ is the conversion efficiency of gravitational energy to stellar radiation,
which is taken to be $0.75$ \citep{Offner+2009}.
From the comparison with the heating rate by the CMB radiation $4\sigma T_\text{rad}^4 \kappa_\text{gr}$, we find that 
within the distance
\begin{align} \label{eq::rcrit}
r_\text{crit} = 1.35 \times 10^3 ~\mathrm{au} \left (\frac{1+z}{21} \right )^{-2}
\left ( \frac{L_\text{acc}}{5.4\times10^{37}~\mathrm{erg~s^{-1}}} \right )^{1/2},
\end{align}
the stellar radiation heating exceeds that by the CMB. 
For example, when $z=20$ and $Z/Z_\odot=10^{-1}$, disk fragmentation occurs at $>10^3~$au away from massive stars (Fig.~\ref{fig_snapshot_m150_zoomin}), where the CMB heating rate is larger than or comparable to that by stellar radiation. 
This indicates that stellar radiation would have smaller impact on fragmentation at such high redshift,
where the CMB heating alone can suppress the formation of low-mass stars with $M_* \lesssim 0.1~M_\odot$. 
The situation is similar when $Z/Z_\odot \gtrsim 10^{-2}$ and $z \gtrsim 10$, where the number of low-mass stars is much smaller than at $z=0$ (Fig.~\ref{fig_mass_spectrum}).
At lower redshifts $z \lesssim 10$, where low-mass stars are abundantly produced by fragmentation and the CMB effect is small, stellar radiation can play some role in suppressing low-mass star formation. 
In fact, equation~\eqref{eq::rcrit} indicates that stellar radiation becomes stronger than the CMB at disk scales ($r\sim \text{a few } 10^3~$au) when $z \lesssim 10$.

Here we study the stellar mass spectra considering the metallicity and redshift as independent parameters. 
In reality, the cloud metallicity should depend on the redshift in a way that the metallicity increases with decreasing redshift due to the accumulation of metals in the ISM and IGM by star formation and supernova explosions. 
The SN explosion of the first stars enriches the pristine gas to the metallicity of $Z/Z_\odot \sim 10^{-6}$--$10^{-3}$
\citep{Greif+2010,Ritter+2015,Smith+2015,Chiaki+2016,Magg+2022}.
The metallicity increases further owing to subsequent star formation and associated SNe.
Several authors have studied metal and dust enrichment owing to star formation at the epoch of the early galaxy formation
by means of cosmological numerical simulations 
\citep[e.g.][]{Wise+2012, Graziani+2015, Graziani+2017, Graziani+2020, Ricotti+2016, Yajima+2017, Jeon+2017, Abe+2021}
as well as semi-analytical calculations \citep[e.g.][]{Inoue2011b, Komiya+2014, Valiante+2016, Sessano+2021}.
Among these, \citet{Wise+2012} have calculated metallicity evolution in a halo with very intense star formation selected from the $\sim$ Mpc simulation box, and found that the typical metallicity increases from $Z \sim 10^{-2}~Z_\odot$ at $z=10$ to $10^{-1}~Z_\odot$ at $z=7$.
Similarly, \citet{Ricotti+2016} have simulated the formation of four galaxies and found that the metallicities of star clusters reach $10^{-1}~Z_\odot$ by $z \lesssim 15$.
\citet{Jeon+2017} have conducted a simulation that follows the chemical evolution of local dwarf galaxies and found that the stellar metallicity becomes $10^{-2}$ -- $10^{-1}~Z_\odot$ at $z=10$--$15$.
Those studies demonstrate that the metallicity can reach values as high as $Z/Z_\odot \sim 0.1$ already at $z \gtrsim 10$, where the CMB can have an important impact on the stellar mass spectrum.
\citet{Bailin+2010} estimated the fraction of stars whose formation might have been affected by CMB heating from numerical simulations of galaxy formation.  
They concluded that about $80\%$ of stars forming at $z \sim 10$ are influenced by CMB radiation, indicating that the CMB has profound impact on star formation in the high-redshift universe.
Our result suggests that SN explosion rate rises by a factor of 1.4 (2.8) at $z=10$ ($z=20$, respectively) 
from the prediction by the Chabrier IMF at a metallicity of $Z/Z_\odot = 0.1$.
Note that numerical simulations show that metal enrichment is highly inhomogeneous and 
there is a large scatter in the gas metallicity in high-$z$ galaxies \citep[e.g.][]{Wise+2012, Smith+2015, Chiaki+2018, Graziani+2020}.
CMB heating makes the stellar mass spectrum at $z \gtrsim 10-20$ more top-heavy than the present-day IMF
even in locally metal-enriched regions, e.g. at the galaxy center.

It is important to consider that most of the numerical studies conducted so far assume a transition in the IMF (from Pop~III to Pop~II/I) at a critical metallicity threshold of $Z/Z_\odot = 10^{-4}$ -- $10^{-3}$, which is at least one order of magnitude smaller than suggested by our results.
With a higher threshold metallicity, massive star formation with a top-heavy IMF would continue longer, yielding more metals in the ISM and IGM in the high-redshift universe. 
In addition, numerical simulations of early galaxy formation
usually adopt a rather small box size of a few Mpc, corresponding to $< 2\sigma$ Gaussian overdensity.
Astrophysically interesting objects, such as high-$z$ quasars, tend to be formed in more biased regions
and their cosmic emergence and star formation take place at earlier epochs.
In fact, the gas-phase metallicity of some high-$z$ quasars is already super-solar before $z=6$
\citep[e.g.][]{Jiang+2007, Juarez+2009, Onoue+2020}.
In studying star formation in such biased regions,
it is important to take into account rapid metal enrichment and thus the enhanced effect of CMB heating. 

Our result that the stellar mass spectrum is biased toward massive stars in a warm-dust environment is also applicable to star formation in dusty starburst galaxies where the intense stellar radiation increases the dust temperature as does the CMB in our case.
The dust temperatures in submillimeter galaxies (SMGs) and ultraluminous infrared galaxies (ULIRGs)
are reported to be $20$--$80$ K \citep[e.g.][]{Chapman+2005,Swinbank+2014,Clements+2018}.
Following observations indirectly show that the stellar IMF in those starburst galaxies is indeed top-heavy.
The number count of the SMGs is larger than expected from the $\Lambda$-CDM model, suggesting larger number of massive stars are formed and thus the luminosity becomes higher at a fixed SFR \citep[e.g.][]{Baugh+2005}.
The abundance ratio of $^{13}$C/$^{18}$O provides another indirect information about the stellar IMF since $^{18}$O ($^{13}$C) is synthesized in stars with $M_* > 8~M_\odot$ ($< 8M_\odot$,  respectively).
\citet{Zhang+2018} found that the line ratio of $^{13}$CO/C$^{18}$O decreases with increasing infrared luminosity $L_\text{IR}$, indicating top-heavy IMF in starburst galaxies such as ULIRGs \citep{Sliwa+2017,Brown+2019} and SMGs \citep{Zhang+2018}.
Our simulation suggests that the dust heating by intense stellar radiation is responsible for the top-heavy IMF in those galaxies
by suppressing the cloud fragmentation.

Metallicity measurement for observed high-$z$ galaxies has seen rapid progress in recent years. 
For example, the \lbrack O~{\sc iii}\rbrack $88 \mu \mathrm{m}$ line 
is detected in galaxies at $z \gtrsim 7$
by Atacama Large Millimeter/submillimeter Array (ALMA) 
\citep{Inoue+2016,Carniani+2017,Laporte+2017, Hashimoto+2018, Tamura+2019}.
Among these, 
\citet{Jones+2020} estimated the gas-phase metallicity for seven galaxies
from the ratio of \lbrack O~{\sc iii}\rbrack $88~\mu \mathrm{m}$ and H$\beta$ luminosities.
The oxygen abundance varies in the range $12 + \log(\mathrm{O/H})=7.6...8.2$, corresponding to $\log \left ( Z/Z_\odot\right ) = -1.4...-0.8$.
Our results suggest that the CMB has impact on star formation in such galaxies leading to a top-heavy IMF.
\citet{Katz+2022} have shown that 
a top-heavy IMF in the early universe can reproduce the observed [C~{\sc ii}]-SFR and [O~{\sc iii}]-SFR relations at $z \gtrsim 6$, by calculating the line-luminosities of simulated galaxies.
The recently launched James Webb Space Telescope (JWST) will 
provide a wealth of information on high-$z$ galaxies, such as the abundance pattern of heavy elements and their ionization state by detecting optical bright lines and will help constrain the environmental conditions where star formation takes place in the early universe.

\section{Summary} \label{sec::conclusion}
We have investigated the impact of the cosmic microwave background radiation (CMB) on the stellar mass spectrum in the early universe,
following the formation of star clusters in various metallicity and redshift environments.
The high-$z$ CMB has a large impact on the evolution of a star-forming cloud and thus on the stellar mass distribution in moderately metal-enriched environments with $Z/Z_\odot \gtrsim 10^{-2}$. 
In the absence of the CMB, 
the filamentary structure generated by collisions of initial turbulent flows promptly fragments and protostars are formed along the filaments, leading to a mass distribution similar to the present-day IMF.
When $z=20$, the temperature floor at $\sim 60~$K set by CMB heating stabilizes the filaments and strongly suppresses fragmentation. 
This reduces the number of low-mass stars and increases the typical fragmentation mass scales, 
leading to a more top-heavy stellar mass spectrum compared to the present-day IMF.

At lower metallicity, when $Z/Z_\odot \lesssim 10^{-3}$, 
the cloud temperature is already higher than the CMB temperature during most of the cloud evolution. 
The CMB has only minor impact on the cloud temperature and thus on the mass distribution of forming stars, which is more top-heavy than the present-day IMF.
When CMB heating is not considered, HD cooling at relatively low densities decreases the gas temperature to $\sim 10$K and therefore the thermal evolution is affected by CMB heating at $z \gtrsim 10$.
The resulting stellar mass spectrum is only marginally affected, but the growth of protostars is accelerated due to the higher gas temperatures in the outer regions of the clouds, which increases the gas accretion rates.
We have analyzed how the typical stellar mass changes with redshift.
Characteristic stellar mass increases monotonically with increasing redshift in mildly metal-enriched environments with $Z/Z_\odot \gtrsim 10^{-2}$, while the redshift evolution is less evident at lower metallicities.
The stellar mass spectra universally consists of a Salpeter-like low-mass component and a massive top-heavy component with a log-flat distribution with the mass fraction in the latter increasing with decreasing metallicity. 
The fraction in the top-heavy component also increases with increasing CMB temperature when $Z/Z_\odot \gtrsim 10^{-2}$ while it hardly changes with redshift at lower metallicity. 

Our result indicates that CMB heating would have significant impact on star formation in high-$z$ galaxies, 
enhancing the supernova rate by a factor of a few at $z=10-20$) from that with the Salpeter-like IMF when $Z/Z_\odot = 0.1$.
Recent numerical simulations have shown that the metallicity in star-forming regions 
reaches $Z/Z_\odot \sim 10^{-2}$ -- $10^{-1}$ already at $z=10$--$15$, 
and we suggest that the resulting stellar mass function in these environments could be more top-heavy than at present.
Observations of high-$z$ galaxies also suggest that $Z/Z_\odot \gtrsim 10^{-1}$ at $z \gtrsim 7$.
Coming observations by James Webb Space Telescope (JWST) will probe star formation in galaxies with $z\gtrsim 10$ and provide an opportunity for understanding the role of CMB heating in the early universe.

\section*{Acknowledgements}
We thank Kazuyki Sugimura, Gen Chiaki, and Hajime Fukushima for fruitful discussion and comments.
This work is financially supported by
the Grants-in-Aid for Basic Research by the Ministry of Education, Science and Culture of Japan 
(SC:19J00324, KO:25287040, 17H01102, 17H02869). 
RS acknowledges support from the Amaldi Research Center funded by the MIUR program Dipartimento di Eccellenza (CUP:B81I18001170001) and funding from the INFN TEONGRAV specific initiative.
We conduct numerical simulation on XC50 at the Center for Computational Astrophysics (CfCA) of the National Astronomical Observatory of Japan and XC40.
We also carry out calculations on XC40 at YITP in Kyoto University.
We use the SPH visualization tool SPLASH \citep{SPLASH} in Figs.~\ref{fig_snapshot}, \ref{fig_snapshot_Tgas} and \ref{fig_snapshot_m150_zoomin}.

\section*{DATA AVAILABILITY}
The data underlying this article will be shared on reasonable request to the corresponding author.

\bibliographystyle{mnras}
\bibliography{biblio2}

\bsp	
\label{lastpage}
\end{document}